\pdfoutput=1

\documentclass[11pt,twoside,a4paper,cmspaper,final,collab]{cms-tdr}
\def\svnVersion{490501P}\def\cmsCernNoTag{CERN-EP-2018-255}\def\cmsCernDate{\today}\def\cmsMessage{Published in the Journal of High Energy Physics as \href{http://dx.doi.org/10.1007/JHEP02(2019)179}{\doi{10.1007/JHEP02(2019)179.}}}

\begin{document}\cmsNoteHeader{EXO-18-001}

\hyphenation{had-ron-i-za-tion}
\hyphenation{cal-or-i-me-ter}
\hyphenation{de-vices}
\RCS$Revision: 488549 $
\RCS$HeadURL: svn+ssh://svn.cern.ch/reps/tdr2/papers/EXO-18-001/trunk/EXO-18-001.tex $
\RCS$Id: EXO-18-001.tex 488549 2019-02-08 14:24:13Z jengbou $
\newlength\cmsFigWidth
\ifthenelse{\boolean{cms@external}}{\setlength\cmsFigWidth{0.85\columnwidth}}{\setlength\cmsFigWidth{0.4\textwidth}}
\ifthenelse{\boolean{cms@external}}{\providecommand{\cmsLeft}{top\xspace}}{\providecommand{\cmsLeft}{left\xspace}}
\ifthenelse{\boolean{cms@external}}{\providecommand{\cmsRight}{bottom\xspace}}{\providecommand{\cmsRight}{right\xspace}}
\providecommand{\CL}{CL\xspace}

\newcommand{\aTriD}{\ensuremath{\alpha_{\mathrm{3D}}}\xspace}
\newcommand{\TriDs}{\ensuremath{D_{\mathrm{N}}}\xspace}
\newcommand{\medip}{\ensuremath{\langle IP_{\mathrm{2D}}\rangle}\xspace}
\newcommand{\pudz}{\ensuremath{PU_{\mathrm{dz}}}\xspace}
\newcommand{\mmed}{\ensuremath{m_{\mathrm{X_{DK}}}}\xspace}
\newcommand{\mdpi}{\ensuremath{m_{\pi_\mathrm{DK}}}\xspace}
\newcommand{\fdpi}{\ensuremath{f_{\pi_\mathrm{DK}}}\xspace}
\newcommand{\taudpi}{\ensuremath{c\tau_{\pi_\mathrm{DK}}}\xspace}
\newcommand{\dmed}{\ensuremath{\mathrm{X}_\mathrm{DK}}\xspace}
\newcommand{\zpv}{\ensuremath{z_{\mathrm{PV}}}\xspace}
\newcommand{\ztrk}{\ensuremath{z_{\text{trk}}}\xspace}
\newcommand{\ipXYs}{\ensuremath{IP_{\text{sig}}}\xspace}
\newcommand{\efb}{\ensuremath{\epsilon_{\text{fb}}}\xspace}
\newcommand{\eflight}{\ensuremath{\epsilon_{\mathrm{fl}}}\xspace}
\newcommand{\efone}{\ensuremath{\epsilon_{\mathrm{f1}}}\xspace}
\newcommand{\eftwo}{\ensuremath{\epsilon_{\mathrm{f2}}}\xspace}
\newcommand{\ef}{\ensuremath{\epsilon_{\mathrm{f}}}\xspace}
\newcommand{\fb}{\ensuremath{f_{\mathrm{b}}}\xspace}
\newcommand{\fbone}{\ensuremath{f_{\mathrm{b1}}}\xspace}
\newcommand{\fbtwo}{\ensuremath{f_{\mathrm{b2}}}\xspace}
\newcommand{\dq}{\ensuremath{\mathrm{Q}_{\mathrm{DK}}}\xspace}
\newcommand{\ncdk}{\ensuremath{N_{C_{\mathrm{DK}}}}\xspace}
\newcommand{\mdown}{\ensuremath{m_{\text{down}}}\xspace}
\newcommand{\nemj}{\ensuremath{n_\mathrm{EMJ}}\xspace}
\newcommand{\nbkg}{\ensuremath{N_{\text{bkg,EMJ}}}\xspace}
\newcommand{\pEMJ}{\ensuremath{P_\mathrm{EMJ}}\xspace}
\newcommand{\fEMJ}{\ensuremath{\varphi(\{\nu\})}\xspace}
\newcommand{\nbtag}{\ensuremath{n_{\text{btag}}}\xspace}
\newcommand{\ncomb}{\ensuremath{n_{\text{comb}}}\xspace}
\newcommand{\mmns}{\ensuremath{\text{mm}}\xspace}
\newcommand{\cmns}{\ensuremath{\text{cm}}\xspace}
\providecommand{\cmsTable}[1]{\resizebox{\textwidth}{!}{#1}}

\cmsNoteHeader{EXO-18-001}
\title{Search for new particles decaying to a jet and an emerging jet}

\date{\today}

\abstract{
A search is performed for events consistent with the pair production of a new heavy particle that acts as a mediator between a dark sector and normal matter, and that decays to a light quark and a new fermion called a dark quark. The search is based on data corresponding to an integrated luminosity of 16.1\fbinv from proton-proton collisions at $\sqrt{s}=13\TeV$ collected by the CMS experiment at the LHC in 2016. The dark quark is charged only under a new quantum-chromodynamics-like force, and forms an ``emerging jet'' via a parton shower, containing long-lived dark hadrons that give rise to displaced vertices when decaying to standard model hadrons. The data are consistent with the expectation from standard model processes. Limits are set at 95\% confidence level excluding dark pion decay lengths between 5 and 225\mm for dark mediators with masses between 400 and 1250\GeV. Decay lengths smaller than 5 and greater than 225\mm are also excluded in the lower part of this mass range. The dependence of the limit on the dark pion mass is weak for masses between 1 and 10\GeV. This analysis is the first dedicated search for the pair production of a new particle that decays to a jet and an emerging jet.
}

\hypersetup{
pdfauthor={CMS Collaboration},
pdftitle={Search for new particles decaying to a jet and an emerging jet},
pdfsubject={CMS},
pdfkeywords={CMS, physics, exotica, discovery, hidden valley, dark fermions}}

\maketitle

\section{Introduction}\label{sec:intro}

Although many astrophysical observations indicate the existence of dark matter~\cite{Young2016},
it has yet to be observed in the laboratory.
While it is possible that dark matter has only gravitational interactions,
many compelling models of new physics contain a dark matter candidate that interacts with quarks.
One class of models includes new, electrically-neutral fermions called ``dark quarks'', \dq,
which are not charged under the forces of the standard model (SM)
but are charged under a new force in the dark sector (``dark QCD'')
that has confining properties similar to quantum chromodynamics (SM QCD)~\cite{petraki, Zurek201491}.
Unlike models based on the popular weakly interacting neutral particle paradigm~\cite{Bertone2005279},
such models naturally explain the observed mass densities of baryonic matter and dark matter~\cite{planck}.

We consider, in particular, the dark QCD model of Bai, Schwaller, Stolarski, and Weiler (BSSW)
that predicts ``emerging jets'' (EMJ)~\cite{BYS,Schwaller2015}.
Emerging jets contain electrically charged SM particles that are consistent with having been created
in the decays of new long-lived neutral particles (dark hadrons), produced in a parton-shower process by dark QCD.
In this model, dark QCD has an $SU(\ncdk)$ symmetry, where \ncdk is the number of dark colors.
The particle content of the model consists of the dark fermions, the dark gluons associated with the force,
and a mediator particle that is charged under both the new dark force and under SM QCD,
thus allowing interactions with quarks.
The dark fermions are bound by the new force into dark hadrons.
These hadrons decay via the mediator to SM hadrons.

\begin{figure}[hbtp]\centering
  {\includegraphics[width=0.4\textwidth]{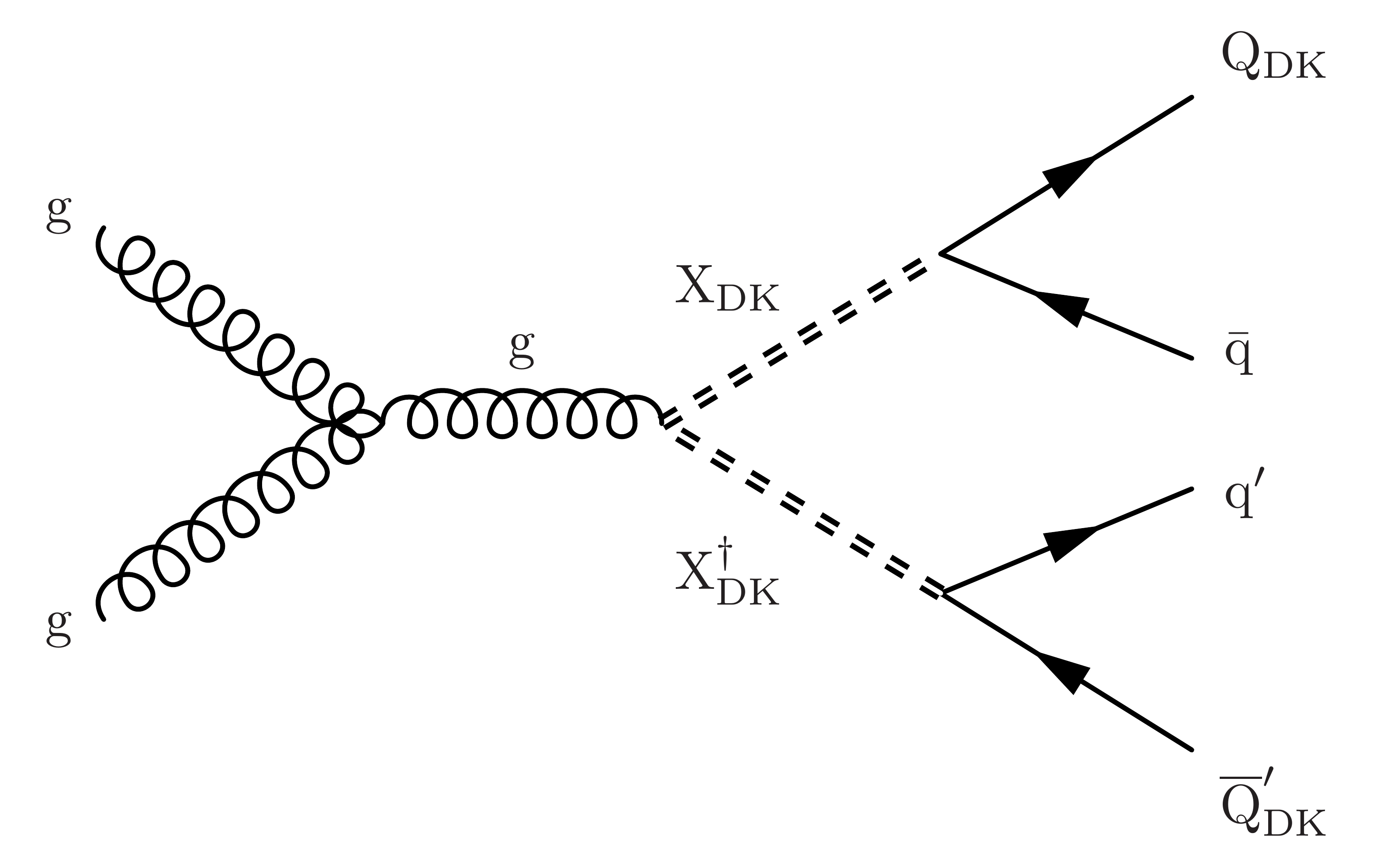}}
  {\includegraphics[width=0.4\textwidth]{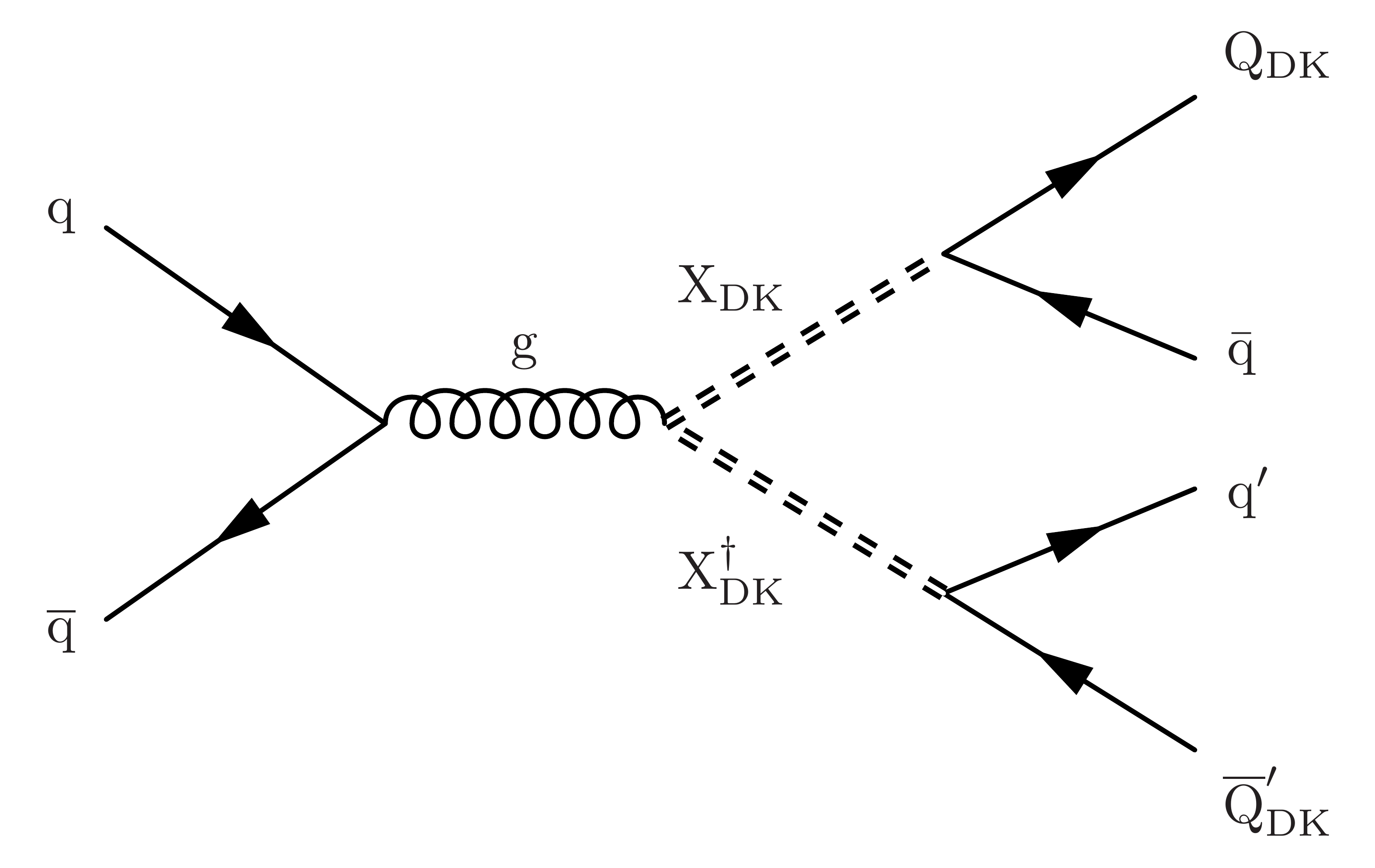}}
  \caption{Feynman diagrams in the BSSW model for the pair production of mediator particles,
    with each mediator decaying to a quark and a dark quark \dq,
    via gluon-gluon fusion (left) and quark-antiquark annihilation (right).}
  \label{fig:DMprod}
\end{figure}

The mediator \dmed is a complex scalar. Under SM QCD, it is an $SU(3)$ color triplet,
and thus can be pair produced via gluon fusion (Fig.~\ref{fig:DMprod}, left)
or quark-antiquark annihilation (Fig.~\ref{fig:DMprod}, right) at the CERN LHC.
The mediator has an electric charge of either ${1}/{3}$ or ${2}/{3}$ of the electron charge,
and it can decay to a right-handed quark with the same charge and a \dq via Yukawa couplings.
There are restrictions on the values of the Yukawa couplings from searches for flavor-changing neutral currents,
neutral meson mixing, and rare decays~\cite{fc1,fc2,fc3,Renner2018}.
We abide by these restrictions by assuming that all the Yukawa couplings are negligible except for the coupling to the down quark~\cite{fc1,fc2,fc3,Renner2018}.

The decay length of the lightest dark meson (dark pion)~\cite{Schwaller2015}, is given by Eq.~\eqref{eqn:bssw}:
\begin{linenomath}
\begin{equation}
c\tau \approx 80\mm
\left( \frac{1}{\kappa^4}\right)
\left( \frac{2\GeV}{\fdpi} \right)^2
\left( \frac{100\MeV}{\mdown} \right)^2
\left( \frac{2\GeV}{\mdpi} \right)
\left( \frac{\mmed}{1\TeV} \right)^4,
\label{eqn:bssw}
\end{equation}
\end{linenomath}
where $\kappa$ is the appropriate element of the $\ncdk{\times}3$ matrix of Yukawa couplings between the mediator particle,
the quarks, and the dark quarks; \fdpi is the dark pion decay constant; and \mdown, \mdpi,
and \mmed are the masses of the down quark, the dark pion, and the mediator particle, respectively.

The signature for this search thus consists of four high transverse momentum (\pt) jets, two from down quarks and two from dark quarks.
The dark quark jets contain many displaced vertices arising from the decays of the dark pions produced in the dark parton shower and fragmentation.
For models with dark hadron decay lengths comparable to the size of the detector,
there can also be significant missing transverse momentum (\ptmiss).
The main background for this signature is SM four-jet production,
where jet(s) are tagged as emerging either because they contain long-lived \PB\ mesons or because of track misreconstruction,
and large artificial \ptmiss is created because of jet energy mismeasurement.
We use a photon+jets data sample to measure the probability for an SM jet to pass selection criteria designed for emerging jets,
and use this probability in estimating the background, as described in Section~\ref{sec:bkgd}.

\section{The CMS detector and event reconstruction}\label{sec:cms}

The CMS detector is a multipurpose apparatus designed to study physics processes in proton-proton ({\Pp\Pp}) and heavy ion collisions.
A superconducting solenoid occupies its central region, providing a magnetic field of 3.8\unit{T} parallel to the beam direction.
The silicon tracker system consists of 1\,440 silicon pixel and 15\,148 silicon strip detector modules.
The trajectories of charged particles within the pseudorapidity range $\abs{\eta}<2.5 $ are reconstructed from
the hits in the silicon tracking system using an iterative procedure with a Kalman filter~\cite{TRK-11-001}.
The tracking efficiency for prompt hadrons is typically over 98\% for tracks with \pt above 1\GeV.
For nonisolated particles with $1<\pt<10\GeV$ and $\abs{\eta}<1.4$,
the track resolutions are typically 1.5\% in \pt and 25--90 (45--150)\mum in the transverse (longitudinal) impact parameter~\cite{TRK-11-001}.
The reconstruction efficiency is low for tracks with an impact parameter larger than 25\cm~\cite{TRK-11-001}.

A lead tungstate crystal electromagnetic calorimeter (ECAL) and
a brass/scintillator hadron calorimeter (HCAL) surround the tracking volume and cover $\abs{\eta}<3$.
A steel and quartz-fiber Cherenkov hadron forward calorimeter extends the coverage to $\abs{\eta}<5$.
The muon system consists of gas-ionization detectors embedded in the steel flux return yoke outside the solenoid,
and covers $\abs{\eta}<2.4$.
The first level of the CMS trigger system~\cite{Khachatryan:2016bia} is designed to select events in less than 4\mus,
using information from the calorimeters and muon detectors.
The high-level trigger (HLT) processor farm then reduces the event rate
to around 1\unit{kHz} before data storage.

A more detailed description of the CMS detector,
together with a definition of the coordinate system and the relevant kinematic variables,
can be found in Ref.~\cite{Chatrchyan:2008zzk}.

The $\Pp\Pp$ interaction vertices are reconstructed by clustering tracks on the basis of their $z$ coordinates along the beamline at their points of
closest approach to the center of the luminous region using a deterministic annealing algorithm~\cite{Rose98deterministicannealing}.
The position of each vertex is estimated with an adaptive vertex fit~\cite{Fruhwirth:2007hz}.
The resolution in the position is around 10--12\mum in each of the three spatial directions~\cite{TRK-11-001}.

The reconstructed vertex with the largest value of summed physics-object $\pt^2$ is taken to be the primary $\Pp\Pp$ interaction vertex (PV).
The physics objects are the jets,
clustered using the jet finding algorithm~\cite{Cacciari:2008gp,Cacciari:2011ma} with the tracks assigned to the vertex as inputs,
and the associated \ptmiss, taken as the negative vector sum of the \pt of those jets.
Other vertices in the same event due to additional {\Pp\Pp} collisions in the same beam crossing are referred to as pileup.

The particle-flow (PF) algorithm~\cite{Sirunyan:2017ulk} is used to reconstruct and identify each individual particle,
with an optimized combination of information from the various elements of the CMS detector.
The energy of each photon is directly obtained from the ECAL measurement, corrected for zero-suppression effects.
The energy of each electron is determined from a combination of the track momentum at the PV,
the corresponding ECAL cluster energy, and the energy sum of all bremsstrahlung photons attached to the track.
The energy of each muon is obtained from the corresponding track momentum.
The energy of each charged hadron is determined from a combination of the track momentum and the corresponding ECAL and HCAL energies,
corrected for zero-suppression effects and for the response functions of the calorimeters to hadronic showers.
Finally, the energy of neutral hadrons is obtained from the corresponding corrected ECAL and HCAL energies.

The analysis involves two types of jets: SM QCD jets and emerging jets. For each event,
the reconstruction of both types of jets starts with the clustering of reconstructed particles with the infrared and collinear safe anti-\kt algorithm~\cite{Cacciari:2008gp,Cacciari:2011ma},
with a distance parameter $R$ of 0.4.
The jet momentum is determined as the vectorial sum of the momenta of associated particles.
Additional identification criteria for the emerging jets are given in Section~\ref{sec:selection}.
For the SM jets, the momentum is found in the simulation to be within 5 to 10\% of the true momentum for jets,
created from the fragmentation of SM quarks and gluons, over the entire \pt spectrum and detector acceptance.
Additional proton-proton interactions within the same or nearby bunch crossings can contribute
additional tracks and calorimetric energy depositions to the jet momentum.
To mitigate this effect,
charged hadrons not associated with the PV are removed from the list of reconstructed particles using
the pileup charged-hadron subtraction algorithm~\cite{Sirunyan:2017ulk},
while an offset correction is applied to correct for remaining contributions~\cite{JetEnCor2011V2,CACCIARI2008119,CMS-PAS-JME-14-001}.
Jet energy corrections are derived from simulation and are confirmed with in situ measurements
with the energy balance of Drell--Yan+jet, dijet, multijet, and photon+jet events~\cite{Khachatryan:2016kdb}.

Jets consistent with the fragmentation of \cPqb\ quarks are identified using
the Combined Secondary Vertex version 2 (CSVv2) discriminator~\cite{BTV-16-002}.
The loose working point corresponds to correctly identifying a \cPqb\ quark jet with
a probability of 81\% and misidentifying a light-flavor jet as a \cPqb\ quark jet with a probability of 8.9\%.

The \ptvecmiss is the negative vector sum of the \ptvec of all PF candidates in an event.
Its magnitude is referred to as \ptmiss.

\section{Simulated samples}\label{sec:mc}
Simulated Monte Carlo (MC) samples are used for the estimation of the signal acceptance A,
defined as the fraction of MC events passing the selection criteria,
and thus including, e.g., tracking and other efficiencies.
These samples are also used for the construction of the templates for background estimation and the validation of background estimation techniques.
The simulation of SM processes, unless otherwise stated,
is performed at leading order in the strong coupling constant using \MGvATNLO 2.2.2~\cite{Alwall:2014hca} or \PYTHIA~8.2~\cite{Sjostrand:2014zea}
with the \textsc{NNPDF3.0}~\cite{Ball:2014uwa} parton distribution functions (PDFs).
The strong coupling constant at the {\cPZ} mass scale is set to 0.130 in the generator.
Parton shower development and hadronization are
simulated with {\PYTHIA} using the underlying-event tune CUETP8M1~\cite{Khachatryan:2015pea}.

Signal samples are generated with the ``hidden valley'' model framework in \PYTHIA~8.212,
using modifications discussed in Ref.~\cite{Schwaller2015}.
The model has several parameters:
the mass of the mediator particle,
the width of the mediator particle,
the number of dark colors,
the number of dark flavors,
the matrix of Yukawa couplings between the \dq and the quarks with the same electric charge as the mediator,
the dark force confinement scale,
the masses of the \dq (one for each dark flavor),
the mass of the dark pion,
the dark pion proper decay length,
and the mass of the dark rho meson.
Following Ref.~\cite{Schwaller2015},
we assume that there are three dark colors and seven dark flavors as suggested in Ref.~\cite{BYS}.
We assume that all \dq (and therefore dark pions) are mass degenerate and that the \dq mass equals the dark force confinement scale.
The mass of the dark pion is assumed to be one half the mass of the \dq.
The mass of the dark rho meson is taken to be four times larger than the mass of the dark pion.
The width of the mediator particle is assumed to be small as compared with the detector mass resolution.
These assumptions leave the mediator mass \mmed, the dark pion mass \mdpi, and the dark pion proper decay length \taudpi as free parameters.
Samples are generated for all permutations of the values of these parameters listed in Table~\ref{tab:sigpar}.
Each set of values defines a single model.

\begin{table}[htb]\centering
  \topcaption{Parameters used in generating the 336 simulated signal event samples.
    A sample corresponding to a single model was created for each possible set of parameter values.}
  \label{tab:sigpar}
  \begin{tabular}{rl}
    \hline
    {Signal model parameters} & \multicolumn{1}{c}{List of values} \\ \hline
    Dark mediator mass \mmed [{\GeVns}]    & \multicolumn{1}{c}{400, 600, 800, 1000, 1250, 1500, 2000} \\
    Dark pion mass \mdpi [{\GeVns}]        & \multicolumn{1}{c}{1, 2, 5, 10}  \\
    Dark pion decay length \taudpi [{\mmns}]    & \multicolumn{1}{c}{1, 2, 5, 25, 45, 60, 100, 150, 225, 300, 500, 1000} \\ \hline
  \end{tabular}
\end{table}

The range in the mediator particle mass over which the search is sensitive depends on the mediator particle pair production cross section.
The mediator particle has the same SM quantum numbers as the supersymmetric partner of an SM quark (squark)~\cite{Schwaller2015}.
Because we assume three dark colors, the signal production cross section is assumed to be three times larger than that for the pair production of a single flavor of squark of the same mass.
We use a calculation of the squark pair production cross section that is based on simplified topologies~\cite{ArkaniHamed:2007fw,Alwall:2008ag,Alwall:2008va,Alves:2011sw,Alves:2011wf}, with other squarks and gluinos decoupled. The cross section is calculated at next-to-leading order in SM QCD with next-to-leading logarithm soft-gluon resummation~\cite{Borschensky:2014cia}.

For all samples, multiple minimum-bias events simulated with \PYTHIA,
with the multiplicity distribution matching that observed in data,
are superimposed with the primary interaction event to model the pileup contribution.
Generated particles are processed through the full \GEANTfour-based simulation of the CMS detector~\cite{GEANT, GEANTdev}.

\section{Event selection}\label{sec:selection}

The analysis is based on data from {\Pp\Pp} collisions at $\sqrt{s}=13\TeV$,
corresponding to an integrated luminosity of 16.1\fbinv collected by the CMS detector in 2016.
The data were obtained using a trigger based on the \pt of the jets in an event.
At the HLT, events were selected if they passed a 900\GeV threshold on the scalar \pt sum of all hadronic jets.
This analysis used only a portion of the data collected during 2016 because, for part of that running period,
saturation-induced dead time was present in the readout of the silicon strip tracker.
Such data were not analyzed because of hard-to-model instantaneous luminosity-dependent inefficiencies for the reconstruction of tracks,
in particular those tracks with impact parameters larger than 10\mm that are key to the selection of the emerging jet signature.

\begin{figure}[hbtp]\centering
  \includegraphics[width=0.55\textwidth]{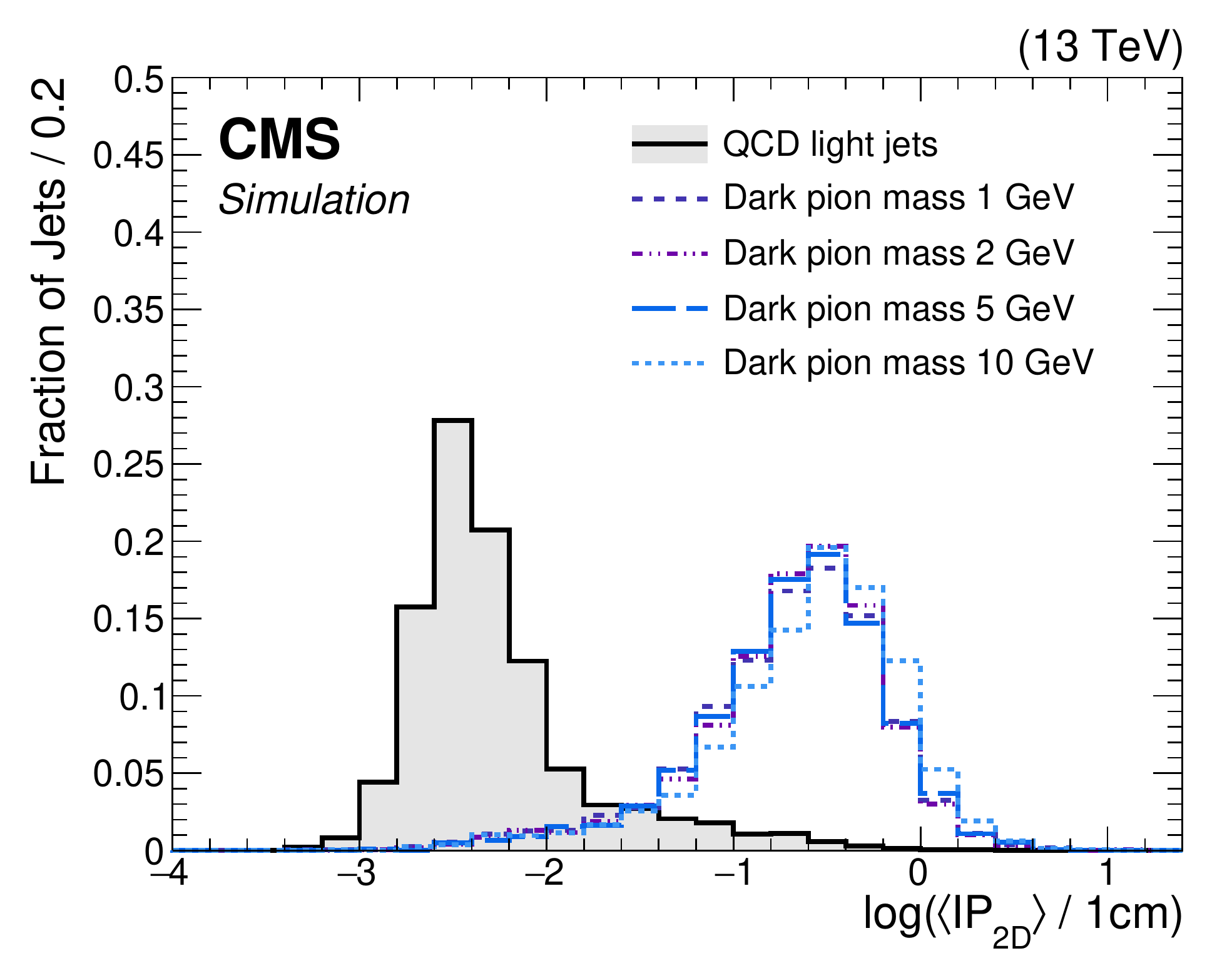}
  \caption{
    Distributions of \medip for background (black) and for signals with a mediator mass of 1\TeV and
    a dark pion proper decay length of 25\mm, for various dark pion masses.}
  \label{fig:meanIP}
\end{figure}

\begin{figure}[hbtp]\centering
  \includegraphics[width=0.55\textwidth]{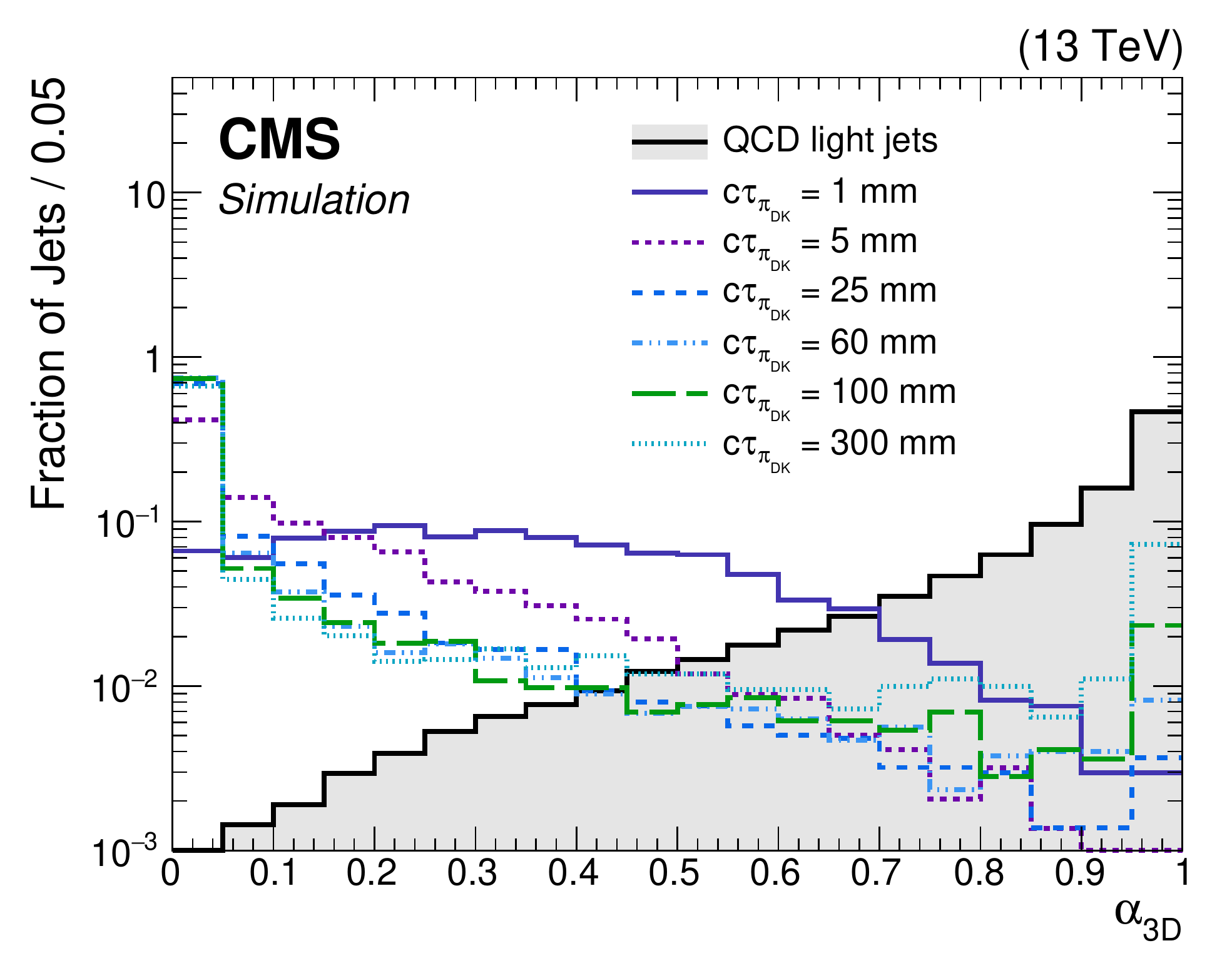}
  \caption{
    Distributions of \aTriD for background (black) and for signals with a mediator mass of 1\TeV and
    a dark pion mass of 5\GeV for dark pion proper decay lengths ranging from 1 to 300\mm.}
  \label{fig:a3D}
\end{figure}

An emerging jet contains multiple displaced vertices and thus multiple tracks with large impact parameters.
Since impact parameter-based variables give good discrimination between SM and emerging jets,
we do not attempt to reconstruct the individual decay vertices of the dark pions.
Emerging jet candidates are required to have $\abs{\eta}<2.0$,
corresponding to the region of the tracker where the impact parameter resolution is best.
Tracks are associated with the candidate if they have $\PT>1\GeV$,
pass the ``high-purity'' quality selection described in Ref.~\cite{TRK-11-001},
and are within a cone of $R=\sqrt{\smash[b]{(\Delta\eta)^2+(\Delta\phi)^2}}=0.4$ (where $\phi$ is azimuthal angle in radians)
around the direction of the jet momentum.
Emerging jet candidates are required to have at least one associated track so that the impact parameter can be estimated.
The jet candidates are also required to have less than 90\% of their energy from electrons and photons, to reduce backgrounds from electrons.
Four variables, similar to the ones defined in Ref.~\cite{EXO-16-003}, are used to select the emerging jets.
The median of the unsigned transverse impact parameters of associated tracks (\medip) is correlated with the dark meson proper decay length,
and should be small for SM jets and large for emerging jets.
The distance between the $z$ position of the track at its distance of closest approach to the PV and the $z$ position of the PV (\pudz) is
used to reject tracks from pileup vertices.
A variable called \TriDs, defined as
\begin{linenomath}
\begin{equation}
\TriDs = \sqrt{ \Big[\frac{\zpv-\ztrk}{0.01\cm}\Big]^2 + [\ipXYs]^2 },
\end{equation}
\end{linenomath}
where \zpv is the $z$ position of the primary vertex,
\ztrk is the $z$ of the track at its closest approach to the PV,
and \ipXYs is the transverse impact parameter significance of the track at its closest approach to the PV,
is used to identify tracks that have an impact parameter that is inconsistent with zero within uncertainties.
The variable \TriDs is smaller for tracks from prompt particles.
A variable called \aTriD,
which is the scalar \pt sum of the associated tracks whose values of \TriDs are smaller than a threshold,
divided by the scalar \pt sum of all associated tracks,
is used to quantify the fraction of the \pt of the jet that is associated with prompt tracks.
This variable should be large for SM jets and small for emerging jets.
Figure~\ref{fig:meanIP} shows the distributions of \medip for background and for signals with a mediator mass of 1\TeV and
a dark pion of various masses and with a proper decay length of 25\mm.
Figure~\ref{fig:a3D} shows the distributions of \aTriD for background and for signals with a mediator mass of 1\TeV and a dark pion mass of 5\GeV.

Since the efficacy of the variables used to select emerging jets depends on the correct identification and reconstruction of the PV,
additional selections are used to remove rare cases observed in simulated background events
where the PV was either not reconstructed or a pileup vertex was chosen as the PV.
We require that the chosen PV be the vertex with the largest scalar \pt sum of its associated tracks.
We also require that the scalar \pt sum of tracks whose extrapolated separation in $z$ from the PV, at the point of closest approach,
is less than 0.01\cm, be larger than 10\% of the sum over all tracks.

Selected candidate events are required to have four jets with $\abs{\eta}<2.0$ and
to pass a threshold on the scalar \pt sum of these jets (\HT).
They must have either two jets tagged as emerging, or one jet tagged as emerging and large \ptmiss.
The selection requirements on the jet-\PT thresholds and the emerging jet selection criteria
were optimized for each signal model listed in Table~\ref{tab:sigpar} as follows.
For each variable listed in Tables~\ref{tab:emjcut} and \ref{tab:cutsets},
a set of potential selection thresholds were chosen based on the distribution of the variable for signal and background.
For each permutation of all the selection thresholds,
we calculated the predicted pseudo-significance for each signal model, defined as $S/\sqrt{S+B+(0.1 B)^2}$,
where $S$ and $B$ correspond to the number of signal and background events
and the $0.1$ corresponds to an estimate of the systematic uncertainty.
In order to limit the final number of background calculations,
the pseudo-significances were used to find the minimum number of selection criteria where
the difference in pseudo-significance between the best selection thresholds and a chosen selection threshold is no more than 10\%,
resulting in a total of seven selection sets.
In Table~\ref{tab:emjcut}, the selection criteria used to select emerging jets are listed.
These jet-level selection criteria, along with event-level kinematic selection criteria,
comprise the final selection criteria, given in Table~\ref{tab:cutsets}.
There are six groups of criteria used to select emerging jets.
The seven selection sets used to define signal regions are given in Table~\ref{tab:cutsets} (sets 1 to 7),
which gives the selections on kinematic variables,
along with the corresponding emerging jet criteria from Table~\ref{tab:emjcut}.
Two basic categories of selections emerge. Other than set 3,
the signal region selection sets require two jets pass emerging jet criteria, and have no requirement on \ptmiss.
Selection set 3 requires that one jet satisfies the emerging jet criteria, and includes a requirement on \ptmiss.
Note that in addition to the \ptmiss requirement,
the EMJ-3 group imposes the loosest criteria on \pudz and \TriDs,
and the tightest requirement on \medip, favoring more displaced tracks.
Selection set 3 is used for signal models with dark pions with large proper decay lengths.
The selection on \medip is large enough that it removes most events containing \cPqb\ quark jets with tracks with
large impact parameters due to the \cPqb\ lifetime;
most SM jets thus selected have tracks with large impact parameters due to misreconstruction.
The substantive requirement on the \ptmiss for this selection set
is essential to attain background rejection equivalent to that obtained when requiring two emerging jet candidates.

Since the initial optimization only used a rough estimate of the systematic uncertainty,
the final selection set for each model is chosen from among the seven as the one that gives the most stringent expected limit,
taking into account more realistic systematic uncertainties.

We also define two additional groups of jet-level criteria that are used to test the effectiveness of the background estimation methods,
described in Section~\ref{sec:bkgd}.
The EMJ-7 group has the same \pudz, \TriDs, and \medip criteria as EMJ-1 set, but loosens only $\aTriD<0.4$,
while the EMJ-8 group has the same \pudz and \TriDs criteria as EMJ-3 set, but loosens $\medip>0.10$ and $\aTriD<0.5$.
These two groups of jet-level criteria are more efficient for quark or gluon jets than those used for the final selections in the analysis,
improving the statistical power of the tests.

\begin{figure}[hbtp]\centering
  \includegraphics[width=0.8\textwidth] {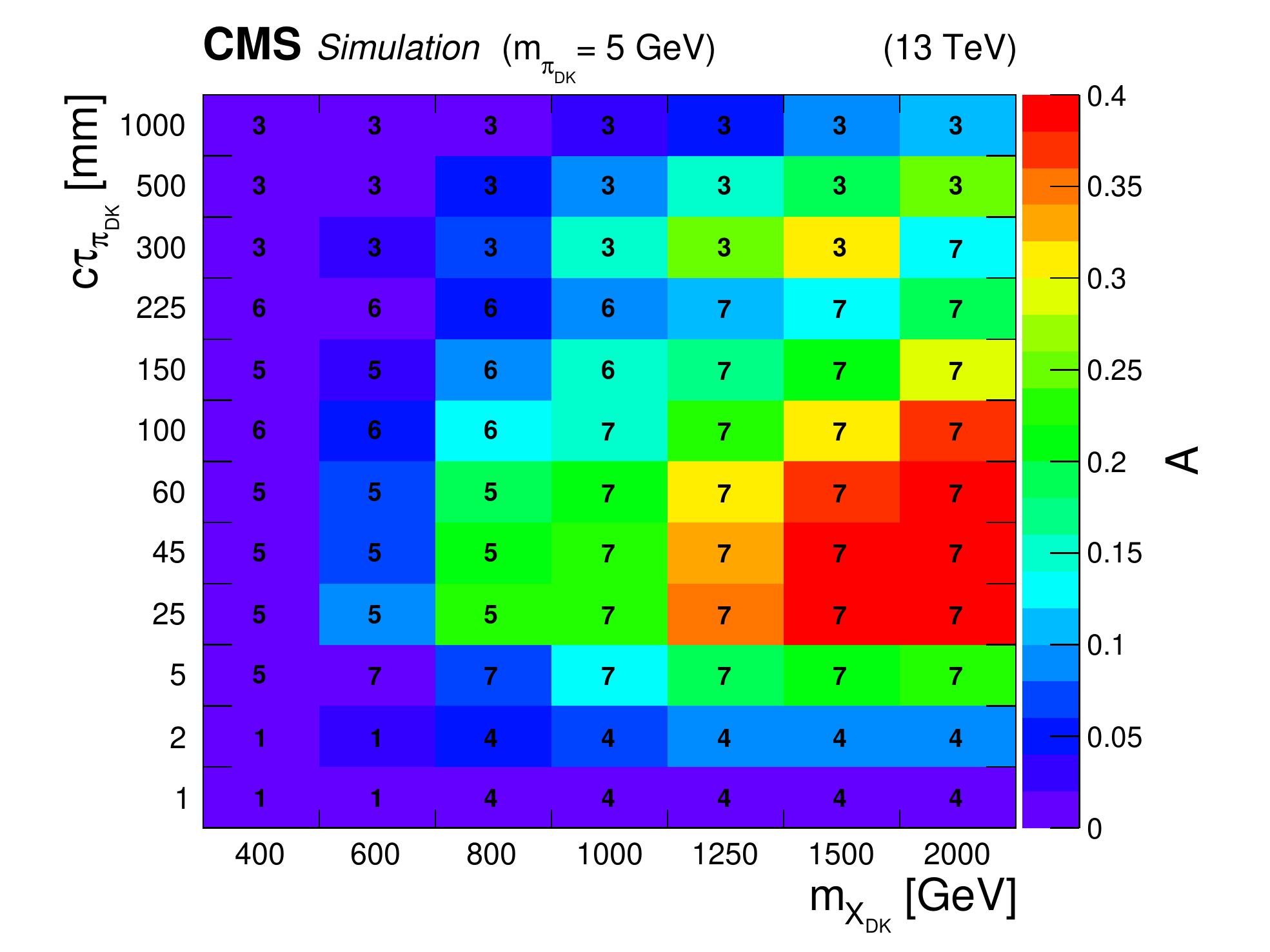}
  \caption{The signal acceptance A,
    defined as the fraction of simulated signal events passing the selection criteria,
    for models with a dark pion mass \mdpi of 5\GeV as a function of
    the mediator mass \mmed and the dark pion proper decay length \taudpi.
    The corresponding selection set number for each model is indicated as text on the plot.}
  \label{fig:sigeff}
\end{figure}

The acceptance of the selection criteria for signal events ranges from a few percent for models with a mediator mass of 400\GeV to 48\% for more massive mediators with a dark pion decay length of 25\mm.
Figure~\ref{fig:sigeff} shows an example of the signal acceptance of models with dark pion mass of 5\GeV as a function of
the mediator mass and the dark pion proper decay length, with text indicating the corresponding selection set number.

\begin{table}[htb]\centering
\topcaption{Groups of requirements (associated operator indicated in parentheses) on the variables used in the identification of emerging jets.
  The groups EMJ-1 to -6 are used for the selection sets that define the signal regions,
  while the groups EMJ-7 and -8 are used to define SM QCD-enhanced samples for the tests of the background estimation methods.
\label{tab:emjcut}}
\begin{tabular}{lcccc}
\hline Criteria group & \pudz ($<$) [{\cmns}]& \TriDs ($<$) & \medip ($>$) [{\cmns}] & \aTriD ($<$)\\ \hline
EMJ-1 &  2.5&    4& 0.05 & 0.25\\
EMJ-2 &  4.0&    4& 0.10 & 0.25\\
EMJ-3 &  4.0&   20& 0.25 & 0.25\\
EMJ-4 &  2.5&    4& 0.10 & 0.25\\
EMJ-5 &  2.5&   20& 0.05 & 0.25\\
EMJ-6 &  2.5&   10& 0.05 & 0.25\\ [-2.ex] \\
EMJ-7 &  2.5&    4& 0.05 & 0.40\\
EMJ-8 &  4.0&   20& 0.10 & 0.50\\
\hline
\end{tabular}
\end{table}

\begin{table}[htb]\centering
\topcaption{The seven optimized selection sets used for this search,
  and the two SM QCD-enhanced selections (sets 8 and 9) used in tests of the background estimation methods.
  The headers of the columns are:
  the scalar \pt sum of the four leading jets (\HT) [{\GeVns}],
  the requirements on the \pt of the jets ($p_\mathrm{T,i}$) [{\GeVns}],
  the requirement on \ptmiss [{\GeVns}],
  the minimum number of the four leading jets that pass the emerging jet selection (\nemj),
  and the EMJ criteria group described in Table~\ref{tab:emjcut}.
  The last column is the total number of models defined in Table~\ref{tab:sigpar} for which
  the associated selection set gives the best expected sensitivity.
\label{tab:cutsets}}
\cmsTable{
\begin{tabular}{cccccccccc}
\hline Set number & \HT
& $p_\mathrm{T,1}$ & $p_\mathrm{T,2}$ & $p_\mathrm{T,3}$
& $p_\mathrm{T,4}$ & $\ptmiss$ & $\nemj (\geq)$
& EMJ group & no. models\\ \hline
1 &  900 & 225 & 100 & 100 & 100 &   0   & 2 &  1 & 12\\
2 &  900 & 225 & 100 & 100 & 100 &   0   & 2 &  2 & 2\\
3 &  900 & 225 & 100 & 100 & 100 & 200 & 1 &  3 & 96\\
4 & 1100 & 275 & 250 & 150 & 150 &   0  & 2 &  1 & 49\\
5 & 1000 & 250 & 150 & 100 & 100 &   0  & 2 &  4 & 41\\
6 & 1000 & 250 & 150 & 100 & 100 &   0  & 2 &  5 & 33\\
7 & 1200 & 300 & 250 & 200 & 150 &   0  & 2 &  6 & 103\\ [-2.ex] \\
8 &  900 & 225 & 100 & 100 & 100 &   0   & 2 &  7 & \multirow{2}{*}{SM QCD-enhanced}\\
9 &  900 & 225 & 100 & 100 & 100 & 200 & 1 &  8 & \\
\hline
\end{tabular}
}
\end{table}

\section{Background estimation}\label{sec:bkgd}

The production of events containing four SM jets can mimic the signal when two of the jets pass the emerging jet criteria,
or when one passes and jet mismeasurement results in artificial \ptmiss.
The background contributions for each of the selection sets are calculated in two different ways,
using the probability for an SM QCD jet to pass the emerging jet requirements.

In the first method, for selection sets 3 and 9 that require at least one emerging jet candidate and \ptmiss,
the background is calculated using Eq.~\eqref{eqn:fakeit2},
\begin{linenomath}
\begin{equation}
\nbkg=\sum_{\mathrm{events}}\pEMJ,
\label{eqn:fakeit2}
\end{equation}
\end{linenomath}
where \nbkg is the predicted background
and \pEMJ is the probability for at least one of the four leading \pt jets to pass the emerging jet criteria.
The sum is over all events in a ``control sample'' defined using all the selection requirements for this set
except for the requirement of at least one emerging jet candidate.
Instead, events are vetoed if one of the four leading \pt jets passes the emerging jet selection.
The misidentification probability of each jet is calculated using Eq.~\eqref{eqn:avgfake}.
\begin{linenomath}
\begin{equation}
\ef = \efb \fb +\eflight \left( 1-\fb \right)
\label{eqn:avgfake}
\end{equation}
\end{linenomath}
Here \efb is the misidentification probability for \cPqb\ jets,
\eflight is the misidentification probability for light-flavor jets,
and \fb is the probability that the jet is a \cPqb\ jet.
The methodology used to estimate \efb, \eflight, and \fb is described below.
The probability \pEMJ is calculated as shown in Eq.~\eqref{eqn:fakeit}.
\begin{linenomath}
\begin{equation}
\begin{split}
\pEMJ&=\sum_{i\in \mathrm{jets}} \ef
\prod_{j\neq i}\left( 1- \ef \right) \\
&+\frac{1}{2}\sum_{i, j\in \mathrm{jets}} \ef \ef
\prod_{k\neq i, j}\left( 1- \ef \right)\\
&+\frac{1}{3}\sum_{i, j, k\in \mathrm{jets}} \ef \ef \ef
\prod_{m\neq i, j, k}\left( 1- \ef \right) + \frac{1}{4}\sum_{i,j,k,m \in \mathrm{jets}}\ef \ef \ef \ef
\end{split}
\label{eqn:fakeit}
\end{equation}
\end{linenomath}
The other selection sets (1 to 8, excluding set 3) require at least two of the
four \pt leading jets to pass emerging jet selection requirements.
The background is estimated using Eq.~\eqref{eqn:fakeit2} as well,
except that the control sample requires exactly one jet to pass the corresponding emerging jet criteria
as well as all other selection requirements for the selection set.
In this case, \pEMJ is the probability for one additional
jet to pass the emerging jet requirements, and is calculated
using Eq.~\eqref{eqn:fakeit3}.
\begin{linenomath}
\begin{equation}
\begin{split}
P_{\mathrm{EMJ}}&=\frac{1}{2}
\sum_{i\in \mathrm{jets\,not\,candidate}} \ef
\prod_{j\neq i}\left( 1- \ef \right)\\
&+\frac{1}{3}
\sum_{i, j\in \mathrm{jets\,not\,candidate}} \ef \ef
\prod_{k\neq i}\left( 1- \ef \right)\\
&+\frac{1}{4}
\sum_{i, j, k\in \mathrm{jets \,not\,candidate}} \ef \ef \ef
\end{split}
\label{eqn:fakeit3}
\end{equation}
\end{linenomath}
In Eq.~\eqref{eqn:fakeit3} the sum is over jets that do not pass the emerging jet selection criteria.

The probability for an SM jet to pass the emerging jet selection criteria
(misidentification) depends on the flavor of the
jet and on the number of tracks associated with the jet.
The probability for a jet initiated by a \cPqb\ quark (\cPqb\ jet) to pass the
selection can be a factor of ten larger than that for a jet
initiated by any other type of parton (light-flavor jet).
For EMJ-3, because of the requirement that \medip be large, the
misidentification probability for \cPqb\ jets and light-flavor jets is similar.
The misidentification probability has a strong dependence on track multiplicity,
ranging from a few percent at low track multiplicities,
to values several orders of magnitude smaller at the highest multiplicities.

\begin{figure}[hbtp]\centering
  \includegraphics[width=0.45\textwidth]{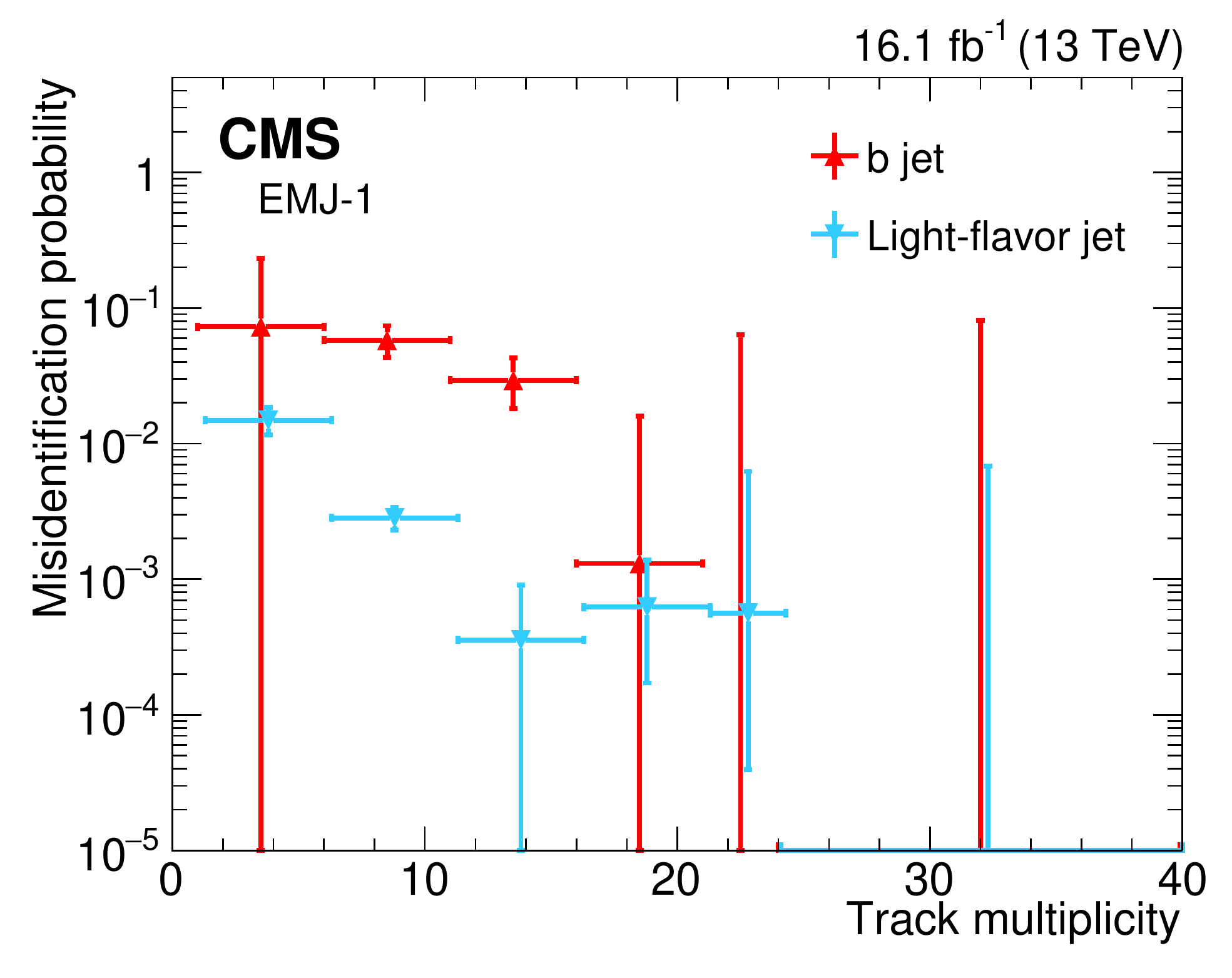}
  \caption{Measured misidentification probability distribution as a function of track multiplicity for
    the EMJ-1 criteria group defined in Table~\ref{tab:emjcut}.
    The red up-pointing triangles are for \cPqb\ jets while the blue down-pointing triangles are for light-flavor jets.
    The horizontal lines on the data points indicate the variable bin width.
    The uncertainty bars represent the statistical uncertainties of \efone, \eftwo, \fbone, and \fbtwo in Eq.~\eqref{eqn:fakerateformula},
    where the uncertainties in \efone and \eftwo correspond to Clopper-Pearson intervals~\cite{ClopperPearson}.}
  \label{fig:misID_cut1a}
\end{figure}

The misidentification probability is measured as a function of track multiplicity using a
sample of events collected with a trigger that requires the presence
of an isolated photon with $\PT>165\GeV$.
We do not expect any signal contamination in this sample.
Two subsamples are created: one with an enhanced and one with a suppressed \cPqb\ quark fraction.
The sample with an enhanced fraction of \cPqb\ jets is selected by requiring
the event to contain at least one additional jet with $\pt>50\GeV$,
beyond the one used in the misidentification probability calculation,
that has a value for the discriminator of the CSVv2 algorithm greater than 0.8.
The sample with suppressed probability of containing a \cPqb\ jet requires an additional jet with $\pt>50\GeV$ with a CSVv2 discriminator value below 0.2.
The \cPqb\ quark fraction of each subsample \fb is determined by fitting the observed distribution of the CSVv2 discriminator to the sum of two templates,
one created using simulated \cPqb\ jets and the other simulated light-flavor jets.
The misidentification probability as a function of the initiating parton type can then be calculated as follows:
\begin{linenomath}
\begin{equation}
\begin{pmatrix} \efb \\ \eflight \end{pmatrix} = \begin{pmatrix} \frac{1-\fbtwo}{\fbone-\fbtwo} & \frac{-(1-\fbone)}{\fbone-\fbtwo} \\ \frac{-\fbtwo}{\fbone-\fbtwo} & \frac{\fbone}{\fbone-\fbtwo}\end{pmatrix}
\begin{pmatrix} {\efone} \\ {\eftwo} \end{pmatrix},
\label{eqn:fakerateformula}
\end{equation}
\end{linenomath}
where \efone, \fbone, \eftwo, and \fbtwo
represent the respective misidentification probability and \cPqb\ jet fraction in the two samples.
Figure~\ref{fig:misID_cut1a} shows the measured misidentification probability for EMJ-1 set.

\begin{figure}[hbtp]\centering
  \includegraphics[width=0.45\textwidth]{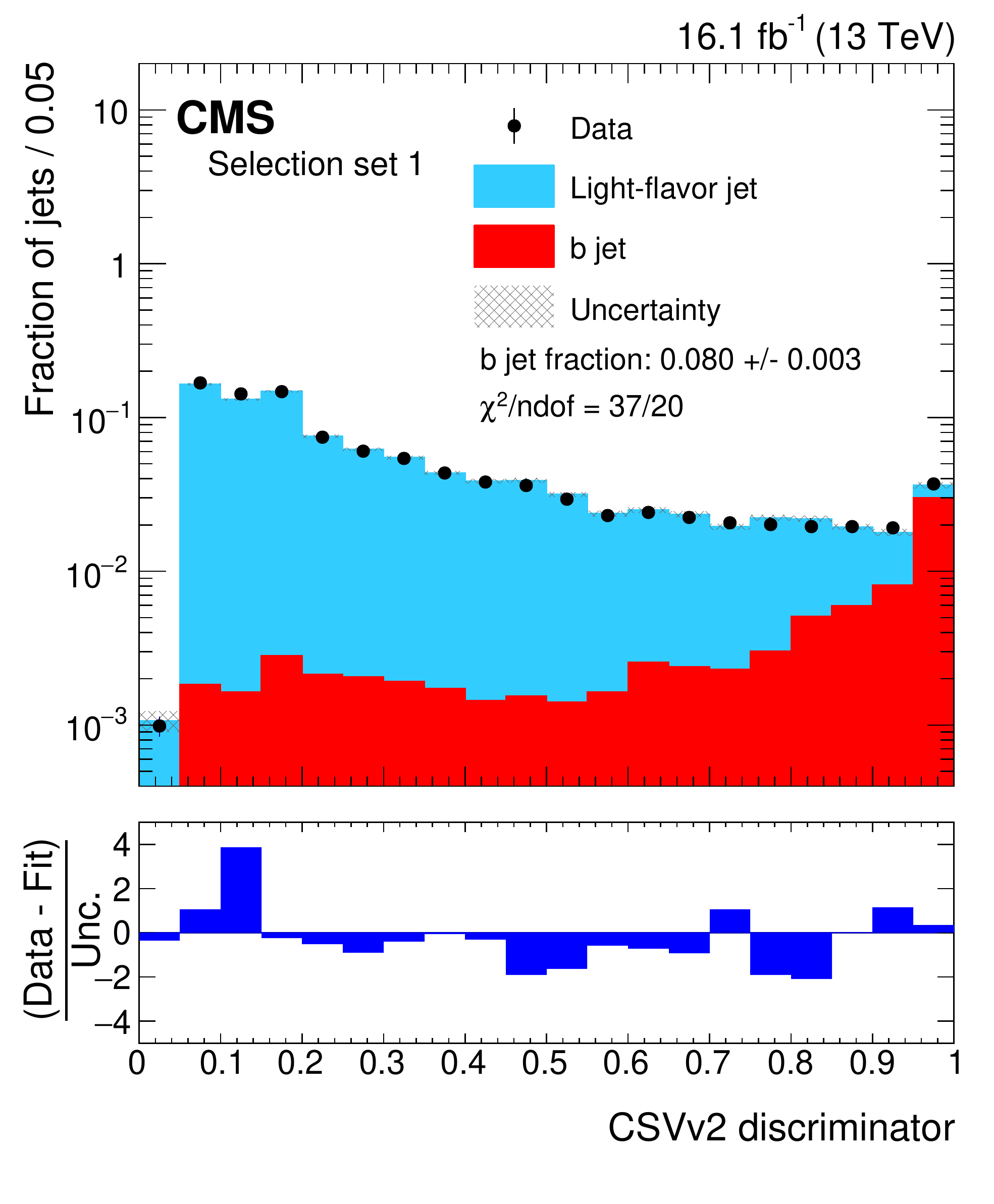}
  \caption{
    Determination of the \cPqb\ jet fraction by fitting the CSVv2 discriminator distribution.
    The red and blue distributions are the CSVv2 discriminator templates of \cPqb\ jets and light-flavor jets, respectively.
    The black points with uncertainty bars show the data distribution.
    The uncertainties in the upper panel include statistical uncertainties of the \cPqb\ jet and light-flavor jet templates,
    and the fit uncertainties, summed in quadrature.
    The goodness of fit is given by the $\chi^2$ divided by the number of degrees of freedom (ndof).
    The bottom panel shows the difference between data and the fit result,
    divided by the combination of the statistical uncertainty of data and the uncertainty from the upper panel.
    The distributions are derived from kinematic samples resulting from selection set 1 in Table~\ref{tab:cutsets}.}
  \label{fig:misID_cut1b}
\end{figure}

When convolving the misidentification probabilities with the kinematic characteristics
and parton composition of the kinematic samples using Eqs.~\eqref{eqn:fakeit} and \eqref{eqn:fakeit3},
the parton composition of the kinematic sample
is determined by fitting the CSVv2 distribution to \cPqb\ jet and light-flavor jet templates obtained from MC simulation.
Figure~\ref{fig:misID_cut1b} shows the resulting fit for the kinematic sample of selection set 1.
The \cPqb\ quark content, \fb, is determined separately for all events and for events with at least one jet passing the emerging jet criteria.
The first is used for predicting the background fraction for selection set 3, which is the only selection set to require only one emerging jet,
the second for the other selection sets.

\begin{figure}[hbtp]\centering
  \includegraphics[width=0.42\textwidth]{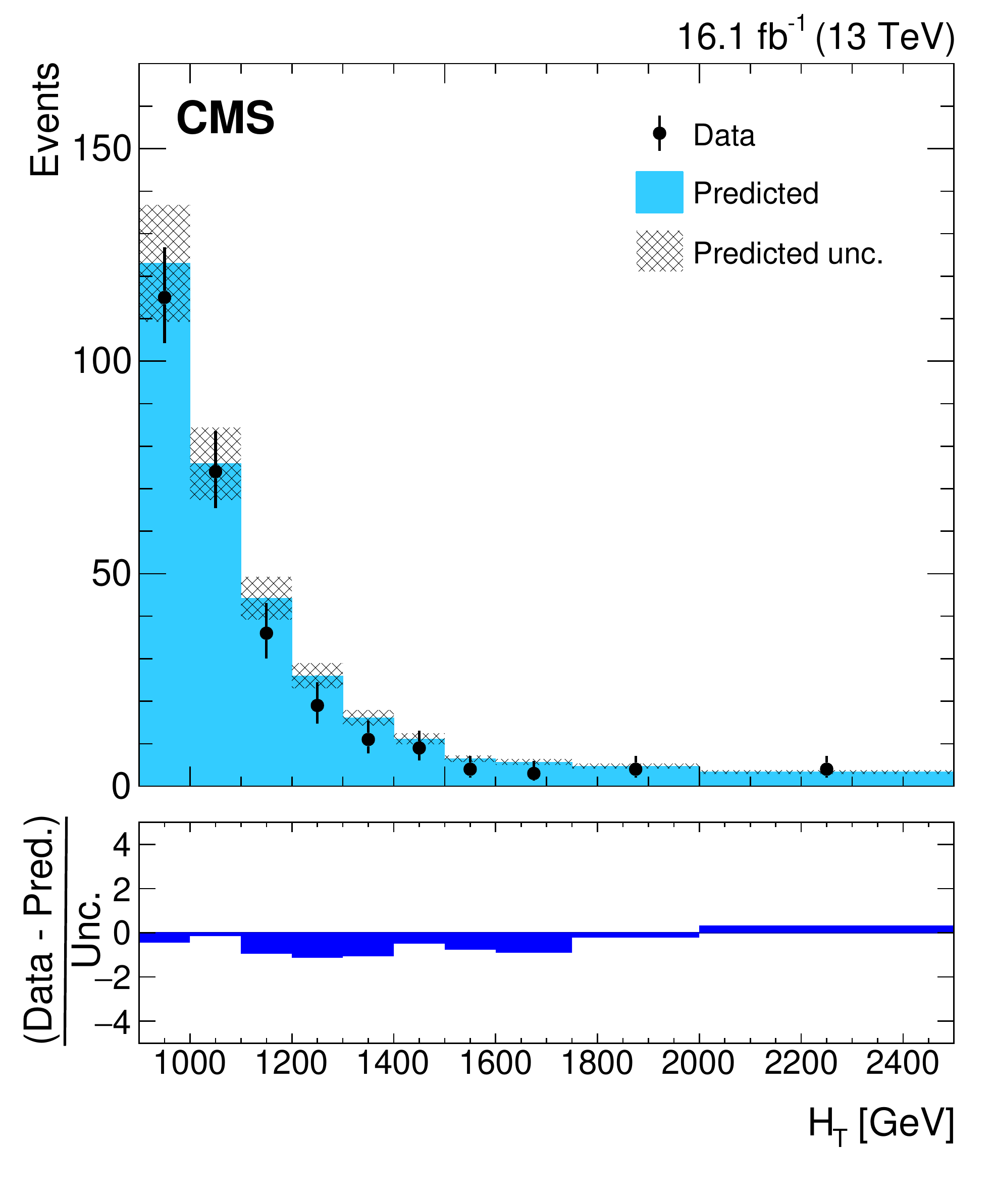}
  \includegraphics[width=0.42\textwidth]{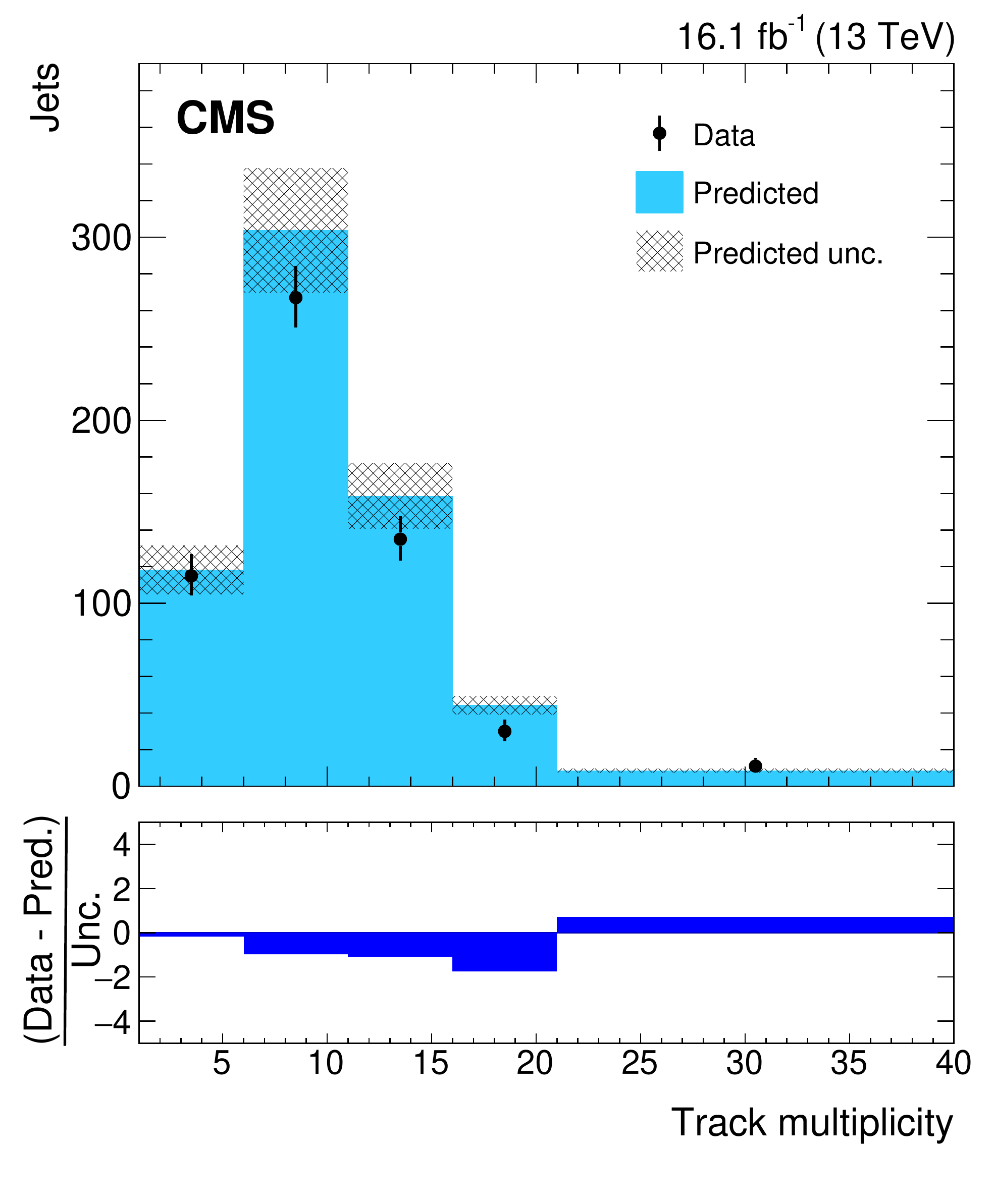}
  \caption{The \HT (left) and number of associated tracks (right)
    distributions for the observed data events (black points) and the predicted background
    estimation (blue) for selection set 8 (SM QCD-enhanced),
    requiring at least two jets tagged by loose emerging jet criteria.
    The bottom panel shows the difference between observed data and predicted background,
    divided by the sum in quadrature of the statistical uncertainty in data and the predicted uncertainties from misidentification probability estimation.}
  \label{fig:data2tag}
\end{figure}

\begin{figure}[hbtp]\centering
  \includegraphics[width=0.42\textwidth]{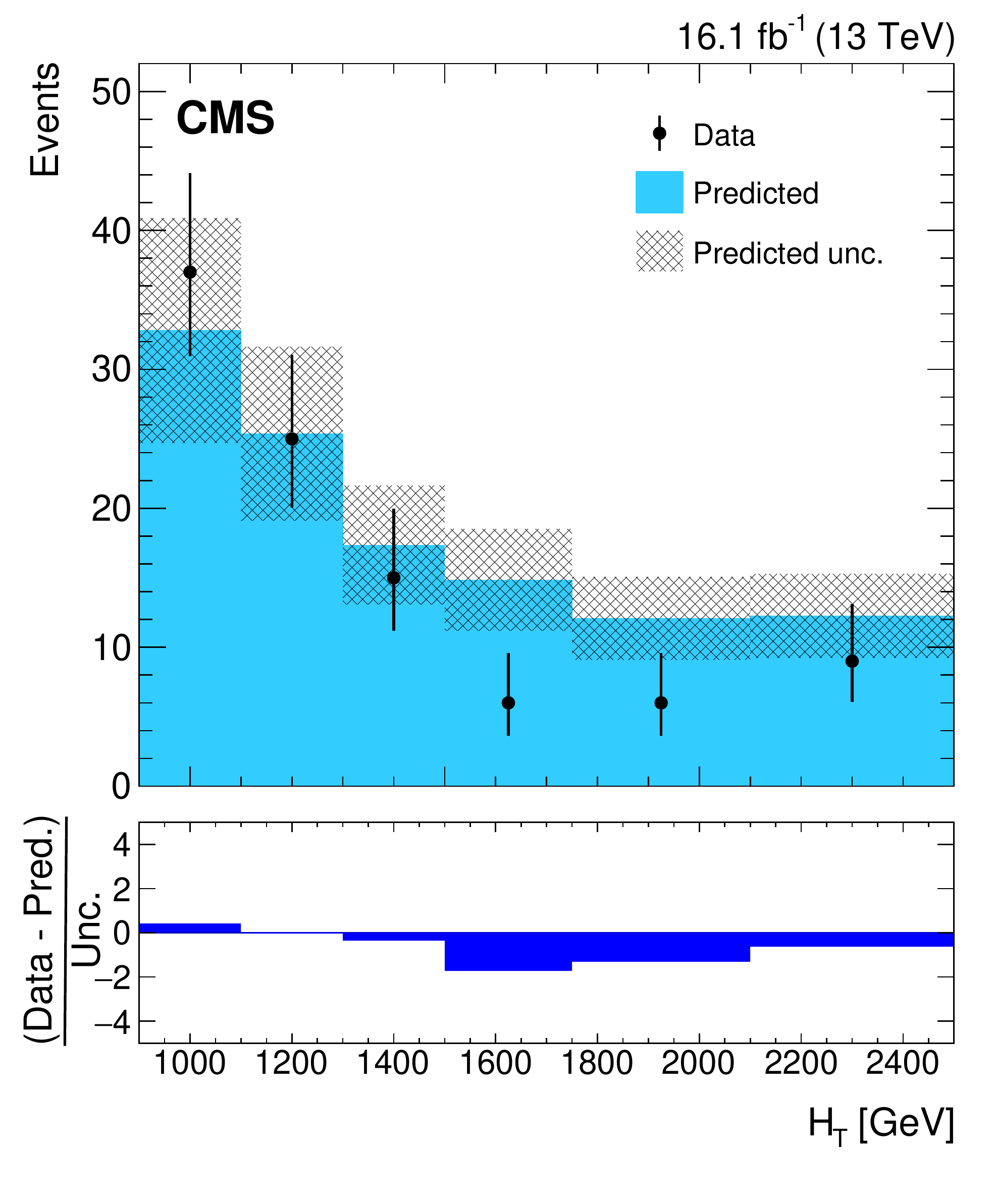}
  \includegraphics[width=0.42\textwidth]{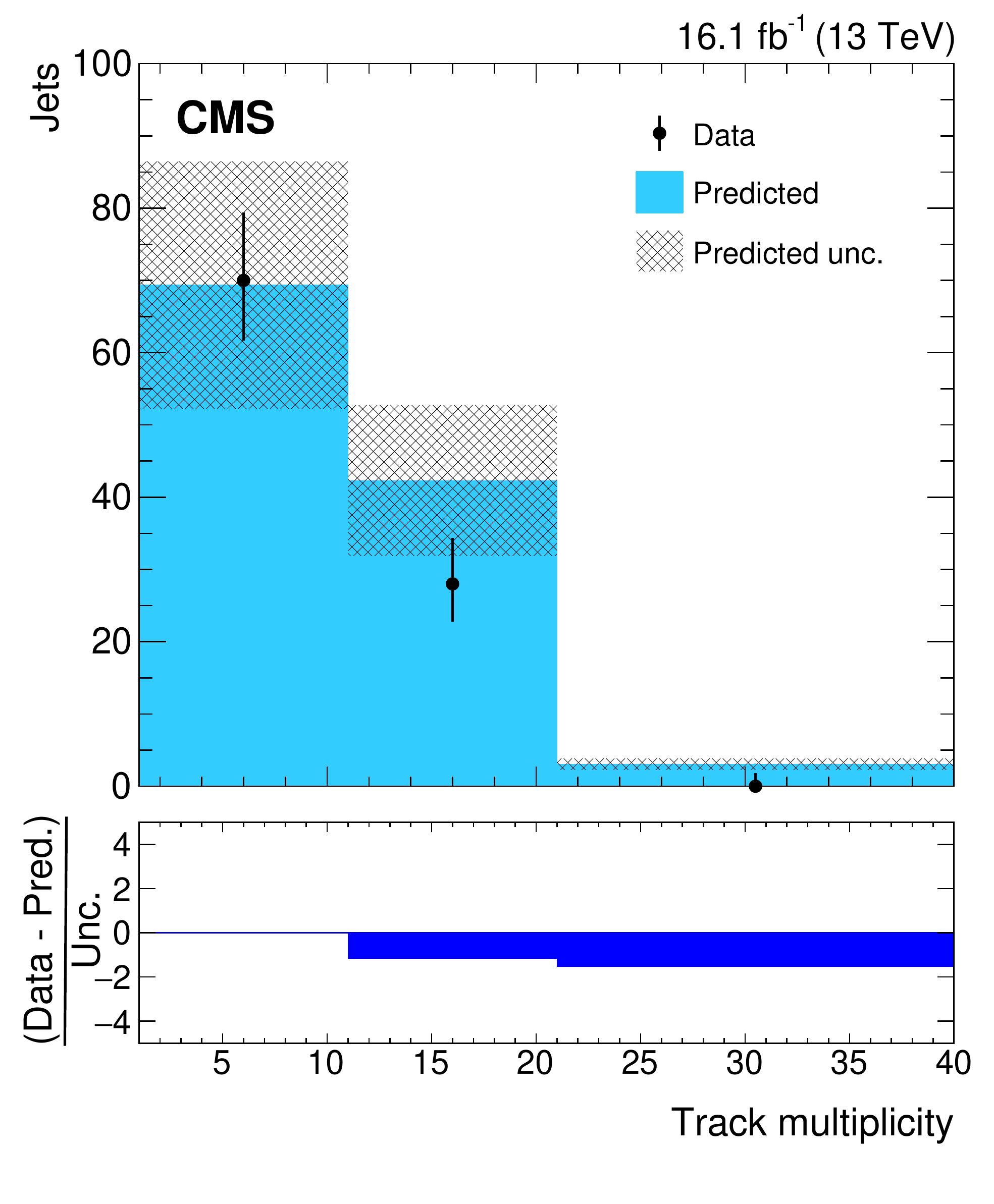}
  \caption{The \HT (left) and number of associated tracks (right)
    distributions of the observed data events (black points) and the predicted background
    estimation (blue) for selection set 9 (SM QCD-enhanced),
    requiring at least one jet tagged by loose emerging jet criteria and large \ptmiss.
    The bottom panel shows the difference between observed data and predicted background,
    divided by the sum in quadrature of the statistical uncertainty in data and the predicted uncertainties from misidentification probability estimation.}
  \label{fig:data1tag}
\end{figure}

The method for estimating the background was tested by using the same procedure on simulated samples,
verifying that the predicted number of selected events was in good agreement with the results obtained when applying the selection criteria to the samples.
For example, the average expected number of events obtained by applying the background estimation method
to simulated samples (average expected number of events passing the selection in simulated samples)
are $207 \pm 30~(231 \pm 18)$
and $52.8 \pm 9.2~(52.1 \pm 6.2)$ for selection sets 8 and 9, respectively.
The background estimation method was also verified using data in the SM QCD-enhanced regions,
and the predicted (observed) numbers of events are $317 \pm 35~(279)$ and $115 \pm 28~(98)$,
as shown in Figs.~\ref{fig:data2tag} and~\ref{fig:data1tag} for selection sets 8 and 9, respectively.
The uncertainty in the predicted number combines those due to the
number of events in the control sample
and statistical uncertainties
in the misidentification probabilities.

The background estimation was also tested using a second method for estimating the fraction of \cPqb\ jets in the control samples.
The distribution of the measured number of \cPqb\ jets (\nbtag) per event in a sample is related to the distribution of the true number of \cPqb\ jets per event,
the distribution of the true number of non-\cPqb\ jets, the identification probability for \cPqb\ jets, and the misidentification probability for non-\cPqb\ jets.
This relationship can be written in the form of a matrix:
\begin{linenomath}
\begin{equation}
\begin{pmatrix}
N_{\mathrm{m,0}} \\
N_{\mathrm{m,1}} \\
N_{\mathrm{m,2}} \\
N_{\mathrm{m,3}} \\
N_{\mathrm{m,4}}
\end{pmatrix}
=
\begin{pmatrix}
A_{\mathrm{0,0}} & A_{\mathrm{0,1}}  &A_{\mathrm{0,2}} & A_{\mathrm{0,3}}& A_{\mathrm{0,4}} \\
A_{\mathrm{1,0}} & A_{\mathrm{1,1}}  &A_{\mathrm{1,2}} & A_{\mathrm{1,3}}& A_{\mathrm{1,4}} \\
A_{\mathrm{2,0}} & A_{\mathrm{2,1}}  &A_{\mathrm{2,2}} & A_{\mathrm{2,3}}& A_{\mathrm{2,4}} \\
A_{\mathrm{3,0}} & A_{\mathrm{3,1}}  &A_{\mathrm{3,2}} & A_{\mathrm{3,3}}& A_{\mathrm{3,4}} \\
A_{\mathrm{4,0}} & A_{\mathrm{4,1}}  &A_{\mathrm{4,2}} & A_{\mathrm{4,3}}& A_{\mathrm{4,4}}
\end{pmatrix}
\begin{pmatrix}
N_{\mathrm{t,0}} \\
N_{\mathrm{t,1}} \\
N_{\mathrm{t,2}} \\
N_{\mathrm{t,3}} \\
N_{\mathrm{t,4}}
\end{pmatrix},
\label{eqn:method2}
\end{equation}
\end{linenomath}
where $N_{\mathrm{t,i}}$ is the number of events with $\mathrm{i}$ \cPqb\ jets and $\mathrm{4-i}$ non-\cPqb\ jets,
$N_{\mathrm{m,i}}$ is the number of events with $\mathrm{i}$ jets passing the CSVv2 loose identification requirements and $\mathrm{4-i}$ failing them,
and $A_{\mathrm{i,j}}$ is the appropriate combination of the CSVv2 efficiencies for a \cPqb\ jet to pass the identification requirement and for a non-\cPqb\ jet to pass the identification requirement, including combinatorics. As these probabilities depend on the jet kinematics, the value used is a weighted sum over the jets in the events.
This matrix can be inverted to get the number of events as a function of true \cPqb\ jet multiplicity from the number of events as a function of the number of identified \cPqb\ jets.
Once the true \cPqb\ jet and non-\cPqb\ jet multiplicities are known,
the misidentification probabilities measured from the photon+jets data can be applied.

To build the matrix, first a sample of events passing all the selection requirements of a selection set,
except the requirement on the number of emerging jet candidates, is selected.
This sample is dominated by SM four-jet production.
The number of events with zero, one, two, three,
or all of the four leading jets satisfying the CSVv2 loose working point is counted,
and the array described in Eq.~\eqref{eqn:method2} is constructed.
The array is inverted to obtain the probability $w(\{\nu\},\nbtag)$
for each of the $\{\nu\}$ possibilities for the true number of \cPqb\ quarks (0--4).
The background is then calculated using Eq.~\eqref{eq:qcdbkg_bu1},
where each probability is weighted with the appropriate combination of misidentification probabilities,
efficiencies, and their combinatorics.
\begin{linenomath}
\begin{equation}
  \nbkg(\nemj)=\sum_{\mathrm{events}}\sum^{4}_{\nu=0} \pEMJ(\nemj|\{\nu|\nbtag\})
  \label{eq:qcdbkg_bu1}
\end{equation}
\end{linenomath}
The probability \pEMJ represents
the probability of having at least \nemj jets pass the emerging jet selections
given $\nu$ true \cPqb\ jets, and
is calculated using Eq.~\eqref{eq:qcdbkg_bu2}.
\begin{linenomath}
\begin{equation}
  \begin{split}
    &\pEMJ(\nemj|\{\nu|\nbtag\})=\sum_{\{\nemj|\{\nu\}\}} \frac{w(\{\nu\},\nbtag)}{\ncomb(\nu)} \prod_{i \in \{\nemj\}}p_{i}\prod_{j \neq i}(1-p_{j})\\
    &p_{k}=p_{k}(\fEMJ)=
    \begin{cases}
      \efb\\
      \eflight
    \end{cases}\\
    &\ncomb(\nu)=\binom{4}{\nu}=\frac{4!}{\nu!(4-\nu)!}
  \end{split}
  \label{eq:qcdbkg_bu2}
\end{equation}
\end{linenomath}
Here $p_{k}$ is the flavor-dependent misidentification probability of jet $k$,
and \fEMJ represents all possible flavor assignments of the four jets.
The combinatoric factor (\ncomb) is the binomial coefficient,
to account for combinatorics in each permutation in $\{\nu\}$.

The respective numbers of predicted background events for selection sets 8 and 9 are $209.2 \pm 1.3 $ and $53.1 \pm 1.2$ in simulated samples,
and are $312.2 \pm 2.0$ and $112.0 \pm 1.6$ for data in SM QCD-enhanced regions.
The predicted numbers include only the uncertainty due to the control sample event statistics.
The predictions are in good agreement with the primary background estimation method.

\section{Systematic uncertainties}\label{sec:syst}

The main sources of systematic uncertainty in the background estimate are
due to the limited number of events in the photon+jets data and in the simulated samples used
for the misidentification probability estimation.
Two other sources are the uncertainties in the determination of \fb for each of the samples used in
the misidentification probability determination and the uncertainties due to differences in the composition of
the non-\cPqb\ jets in the sample used in determining the misidentification probability compared to that in the kinematic samples.
We estimate the first uncertainty by using the value of \fb predicted by simulation instead of that obtained by the template fit.
We estimate the second uncertainty by using the method on MC simulation.
The uncertainty is estimated as
the difference in the prediction when
using a misidentification probability determined using
an MC sample of events containing a high-\PT photon
and when using a misidentification probability
determined using an MC sample of SM QCD multijet production.
The estimated resulting uncertainty for each selection set is given in Table~\ref{tab:bckuncertainties}.

\newcolumntype{C}[1]{>{\centering\arraybackslash}p{#1}}
\newcolumntype{d}{D{.}{.}{-1} }
\begin{table}[htb]
  \topcaption{
    Systematic uncertainties affecting the background estimate from control samples in data.
    For the definition of the selection sets, see Table~\ref{tab:cutsets}.
  }
  \label{tab:bckuncertainties}
  \centering
  \begin{tabular}{cC{4.5cm}d}
    \hline
    \multirow{2}{*}{Set number} & \multicolumn{2}{c}{Source of uncertainty (\%)} \\
    & \cPqb\ quark fraction & \multicolumn{1}{c}{non-\cPqb\ quark composition} \\ \hline
    1 & 2.8 & 1.4 \\
    2 & 0.6 & 4.4 \\
    3 & 2.9 & 28.3 \\
    4 & 5.0 & 4.4 \\
    5 & 0.9 & 4.0 \\
    6 & 1.6 & 2.1 \\
    7 & 1.0 & 6.3 \\
    \hline
  \end{tabular}
\end{table}

The main source of uncertainty in the estimation of the signal acceptance is the modeling of displaced tracks in the simulation.
Other sources include uncertainties in
PDFs,
MC modeling of the trigger efficiency,
integrated luminosity determination,
jet energy scale (JES),
pileup reweighting,
and statistical uncertainties due to the limited size of the MC samples.
Systematic uncertainties are largest for the models with the shortest decay lengths.

The uncertainty due to the track modeling in simulation is evaluated by smearing the tracks in signal events
using the resolution functions that respectively transform the simulated distributions of $\zpv-\ztrk$
and 2D impact parameter in photon+jet MC samples so that they agree with those in data.
The change in signal acceptance when using this transformation is taken as the uncertainty.

The acceptance is evaluated using both the MC trigger selection and
using a trigger efficiency determined using SM QCD multijet events.
The difference is taken as an uncertainty in the acceptance.

The uncertainty in the integrated luminosity determination is 2.5\%~\cite{cms_lumi}.
The uncertainty due to pileup modeling is measured by varying the total inelastic cross section by 4.6\%~\cite{Sirunyan2018} and
reweighting the simulation accordingly.
The effect of the JES uncertainty is evaluated by shifting the \pt of jets by the JES uncertainty,
and measuring its effect on signal acceptance~\cite{Khachatryan:2016kdb}.
The shift in signal acceptance is taken as the uncertainty.
We account for variations of the acceptance due to the PDF uncertainties
following the PDF4LHC prescription~\cite{Butterworth2016}.
The resulting ranges of the systematic uncertainties are given in Table~\ref{tab:signcertainties}.

\begin{table}[htb]
  \topcaption{
    Ranges of systematic uncertainties over all models given in Table~\ref{tab:sigpar}
    for which a 95\% \CL exclusion is expected, for the uncertainties from different sources.
  }
  \label{tab:signcertainties}
  \centering
  \begin{tabular}{l r@{\hspace{1.7em}}r@{ -- }ll}
    \hline
    Source          & \multicolumn{4}{c}{Uncertainty (\%)} \\
    \hline
    Track modeling         & & {\textless}1 & 3 \\
    MC event count         & & 2 & 17 \\
    Integrated luminosity  & & \multicolumn{2}{c}{2.5} \\
    Pileup                 & & {\textless}1 & 5 \\
    Trigger                & & 6 & 12  \\
    JES                    & & {\textless}1 & 9 \\
    PDF                    & & {\textless}1 & 4 \\
    \hline
  \end{tabular}
\end{table}

\section{Results}\label{sec:results}

\begin{figure}[hbtp]\vspace{1em}\centering
  {\includegraphics[width=0.49\textwidth]{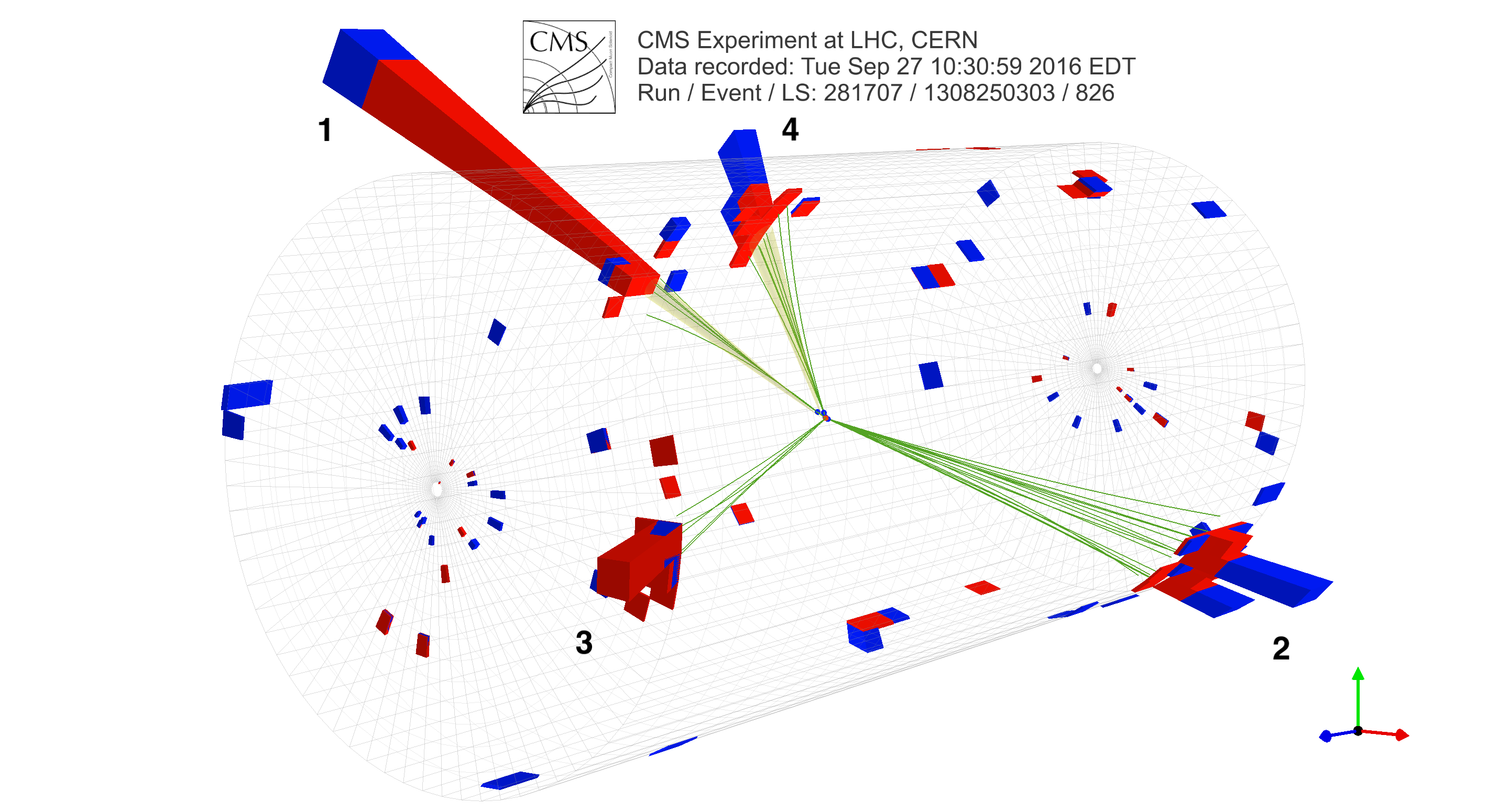}}
  {\includegraphics[width=0.49\textwidth]{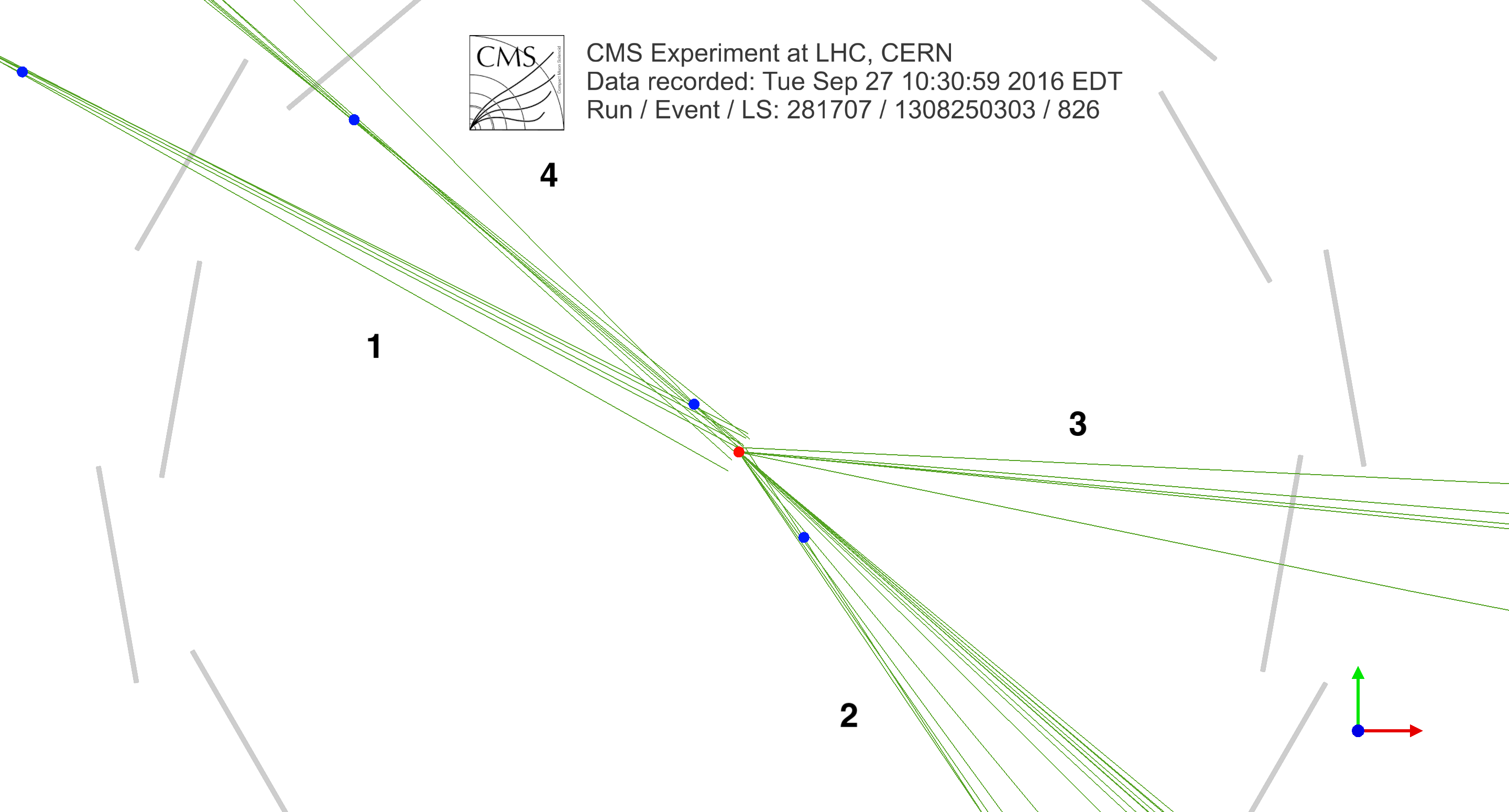}}
  \vspace{1em}
  \caption{
    Event display of an event passing both selection set 1 and selection set 5.
    The event contains four jets (jets 1 and 4 pass the emerging jet criteria),
    consistent with the decay of two massive mediator particles,
    each decaying to an SM quark and a dark QCD quark. In such a scenario,
    the dark mesons produced in the fragmentation of the dark quark would decay back to SM particles via the mediator,
    resulting in displaced vertices with decay distances on the mm scale.
    (Left) 3D display: the green lines represent reconstructed tracks,
    the red (blue) truncated pyramids represent energy in the ECAL (HCAL) detectors, respectively.
    (Right) Reconstructed tracks in $\rho$--$\phi$ view.
    The filled blue circles represent reconstructed secondary vertices, while the filled red circle is the PV.
    The solid grey lines represent the innermost layer of the silicon pixel detector.
  }
  \label{fig:eventdisplay}
\end{figure}

The number of events passing each selection set, along with the background expectation, is given in Table~\ref{tab:results}.
Figure~\ref{fig:eventdisplay} shows a graphical representation of one of the events passing the selection requirements.
This event passes both selection set 1 and selection set 5.
The display on the left shows the four jets.
The display on the right shows the reconstructed tracks in the $\rho$--$\phi$ view.
The filled circles represent reconstructed secondary vertices, while the grey lines represent the innermost layer of the silicon pixel tracker.

\newcolumntype{e}{D{.}{.}{+2} }
\newcolumntype{f}{D{.}{.}{+3} }
\newcolumntype{g}{D{.}{.}{+4} }
\begin{table}[htb]
\centering
\topcaption{Expected ($\mathrm{mean} \pm \mathrm{syst}_1 \pm \mathrm{syst}_2$) and observed event yields for each selection set.
Uncertainties due to the limited number of events in the
control sample and statistical uncertainties in the misidentification probabilities are denoted by ``syst$_1$'',
while ``syst$_2$'' combines the systematic uncertainty sources discussed in Table~\ref{tab:bckuncertainties}.
The ``Signal'' column shows the expected event yield for the heaviest mediator mass that can be excluded for each set,
with the systematic uncertainties from sources discussed in Table~\ref{tab:signcertainties} summed in quadrature.
The associated model parameters are specified in the last three columns.
}
\label{tab:results}
\cmsTable{
\begin{tabular}{ c r@{ }c@{ }r@{ }c@{ }r e r@{ }r@{ }c@{ }r@{ }r egf}
\hline
\multirow{2}{*}{Set number} & \multicolumn{5}{c}{\multirow{2}{*}{Expected}} & \multicolumn{1}{c}{\multirow{2}{*}{Observed}} & & \multicolumn{3}{c}{\multirow{2}{*}{Signal}} & & \multicolumn{3}{c}{Model parameters}\\
& & & & & & & & & & & & \multicolumn{1}{c}{\mmed [{\GeVns}]} & \multicolumn{1}{c}{\mdpi [{\GeVns}]} & \multicolumn{1}{c}{\taudpi [{\mmns}]} \\
\hline
1 &   168 & $\pm$ &   15 & $\pm$ &    5 & 131 & & 36.7 & $\pm$ & 4.0 & & 600 & 5 & 1 \\
2 &  31.8 & $\pm$ &  5.0 & $\pm$ &  1.4 &  47 & & (\,14.6 & $\pm$ & 2.6 & )$\times 10^2$ & 400 & 1 & 60 \\
3 &  19.4 & $\pm$ &  7.0 & $\pm$ &  5.5 &  20 & & 15.6 & $\pm$ & 1.6 & & 1250 & 1 & 150 \\
4 &  22.5 & $\pm$ &  2.5 & $\pm$ &  1.5 &  16 & & 15.1 & $\pm$ & 2.0 & & 1000 & 1 & 2 \\
5 &  13.9 & $\pm$ &  1.9 & $\pm$ &  0.6 &  14 & &  35.3 & $\pm$ & 4.0 & & 1000 & 2 & 150 \\
6 &   9.4 & $\pm$ &  2.0 & $\pm$ &  0.3 &  11 & & 20.7 & $\pm$ & 2.5 & & 1000 & 10 & 300 \\
7 &  4.40 & $\pm$ & 0.84 & $\pm$ & 0.28 &   2 & & 5.61 & $\pm$ & 0.64 & & 1250 & 5 & 225 \\
\hline
\end{tabular}
}
\end{table}

\begin{figure}[hbtp]\centering
  \includegraphics[width=0.8\textwidth] {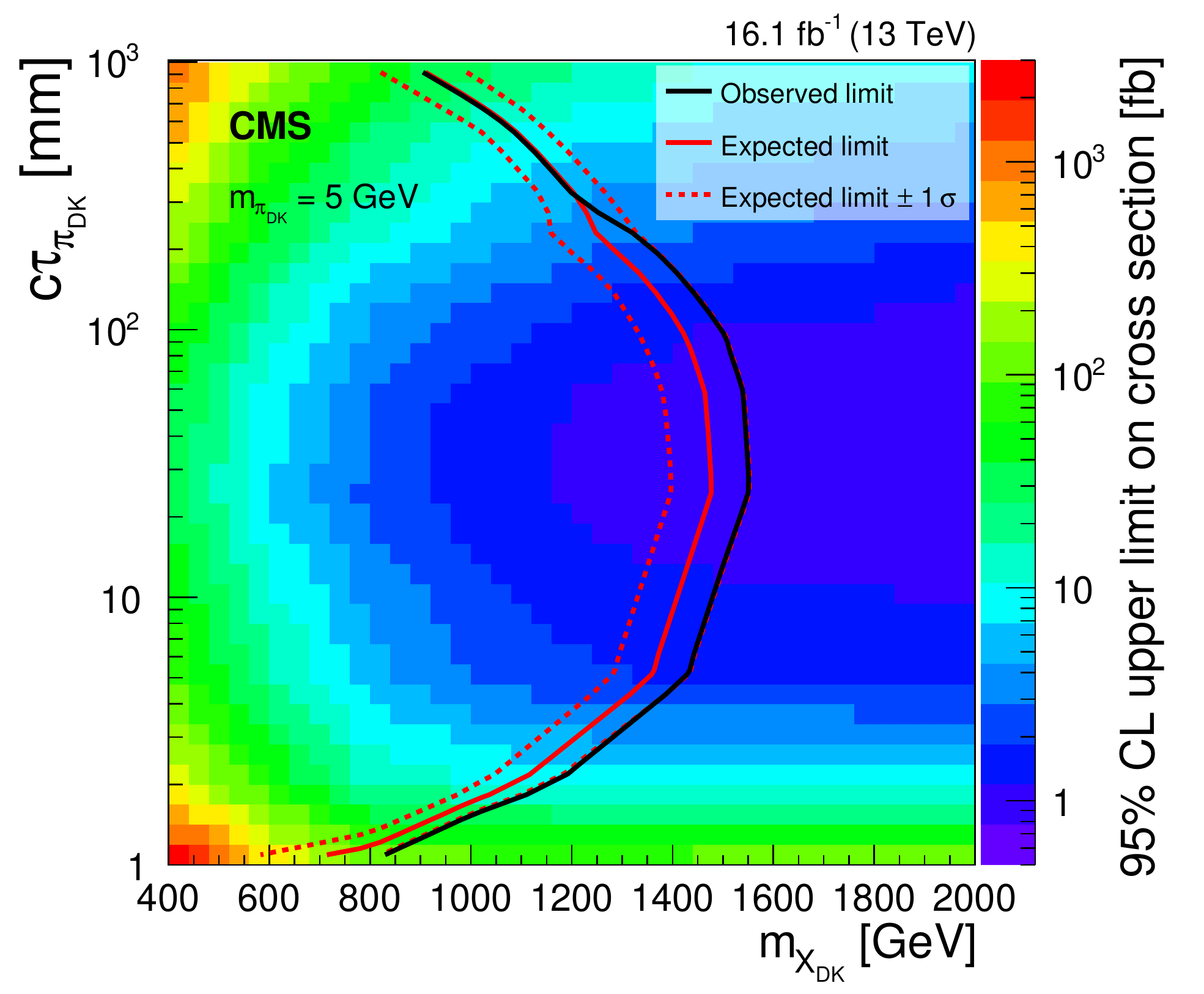}
  \caption{Upper limits at 95\% \CL on the signal cross section and signal exclusion contours derived from theoretical cross sections
    for models with dark pion mass \mdpi of 5\GeV in the $\mmed-\taudpi$ plane.
    The solid red contour is the expected upper limit, with its one standard-deviation region enclosed in red dashed lines.
    The solid black contour is the observed upper limit.
    The region to the left of the observed contour is excluded.
  }
  \label{fig:limitcurve}
\end{figure}

No significant excess with respect to the SM prediction is observed.
A 95\% confidence level (\CL) cross section upper bound is calculated following
the modified frequentist \CLs prescription~\cite{Junk:1999kv,Read:2002hq,CMS-NOTE-2011-005},
using an asymptotic approximation~\cite{Cowan:2010js} for the profile likelihood ratio based test statistic,
where the systematic uncertainties are taken as nuisance parameters.
The 95\% \CL limits on the signal cross section, expected,
and observed exclusion contours on signal parameters are shown in Fig.~\ref{fig:limitcurve} for $\mdpi=5\GeV$.
The dependence of the limit on \mdpi is weak for \mdpi between 1 and 10\GeV.
Dark pion decay lengths between 5 and 225\mm are excluded at 95\% \CL for dark mediator masses between 400 and 1250\GeV.
Decay lengths smaller than 5 and greater than 225\mm are also excluded in the lower part of this mass range.

\section{Summary}\label{sec:summary}
A search is presented for events consistent with the pair production of a heavy mediator particle that decays to a light quark and a new fermion called a dark quark, using data from proton-proton collisions at $\sqrt{s}=13\TeV$ corresponding to an integrated luminosity of 16.1\fbinv.
The dark quark is assumed to be charged only under a new quantum-chromodynamics-like dark force, and to form an emerging jet via a parton shower,
containing long-lived dark hadrons that give rise to displaced vertices when decaying to standard model hadrons.
The data are consistent with the expected contributions from standard model processes.
Limits are set at 95\% confidence level excluding dark pion decay lengths between 5 and 225\mm for dark mediators with masses between 400 and 1250\GeV.
Decay lengths smaller than 5 and greater than 225\mm are also excluded in the lower part of this mass range.
The dependence of the limit on the dark pion mass is weak for masses between 1 and 10\GeV.
This analysis is the first dedicated search for the pair production of a new particle that decays to a jet and an emerging jet.

\begin{acknowledgments}
We congratulate our colleagues in the CERN accelerator departments for the excellent performance of the LHC and thank the technical and administrative staffs at CERN and at other CMS institutes for their contributions to the success of the CMS effort. In addition, we gratefully acknowledge the computing centers and personnel of the Worldwide LHC Computing Grid for delivering so effectively the computing infrastructure essential to our analyses. Finally, we acknowledge the enduring support for the construction and operation of the LHC and the CMS detector provided by the following funding agencies: BMBWF and FWF (Austria); FNRS and FWO (Belgium); CNPq, CAPES, FAPERJ, FAPERGS, and FAPESP (Brazil); MES (Bulgaria); CERN; CAS, MoST, and NSFC (China); COLCIENCIAS (Colombia); MSES and CSF (Croatia); RPF (Cyprus); SENESCYT (Ecuador); MoER, ERC IUT, and ERDF (Estonia); Academy of Finland, MEC, and HIP (Finland); CEA and CNRS/IN2P3 (France); BMBF, DFG, and HGF (Germany); GSRT (Greece); NKFIA (Hungary); DAE and DST (India); IPM (Iran); SFI (Ireland); INFN (Italy); MSIP and NRF (Republic of Korea); MES (Latvia); LAS (Lithuania); MOE and UM (Malaysia); BUAP, CINVESTAV, CONACYT, LNS, SEP, and UASLP-FAI (Mexico); MOS (Montenegro); MBIE (New Zealand); PAEC (Pakistan); MSHE and NSC (Poland); FCT (Portugal); JINR (Dubna); MON, RosAtom, RAS, RFBR, and NRC KI (Russia); MESTD (Serbia); SEIDI, CPAN, PCTI, and FEDER (Spain); MOSTR (Sri Lanka); Swiss Funding Agencies (Switzerland); MST (Taipei); ThEPCenter, IPST, STAR, and NSTDA (Thailand); TUBITAK and TAEK (Turkey); NASU and SFFR (Ukraine); STFC (United Kingdom); DOE and NSF (USA).

\hyphenation{Rachada-pisek} Individuals have received support from the Marie-Curie program and the European Research Council and Horizon 2020 Grant, contract No. 675440 (European Union); the Leventis Foundation; the A. P. Sloan Foundation; the Alexander von Humboldt Foundation; the Belgian Federal Science Policy Office; the Fonds pour la Formation \`a la Recherche dans l'Industrie et dans l'Agriculture (FRIA-Belgium); the Agentschap voor Innovatie door Wetenschap en Technologie (IWT-Belgium); the F.R.S.-FNRS and FWO (Belgium) under the ``Excellence of Science - EOS" - be.h project n. 30820817; the Ministry of Education, Youth and Sports (MEYS) of the Czech Republic; the Lend\"ulet (``Momentum") Program and the J\'anos Bolyai Research Scholarship of the Hungarian Academy of Sciences, the New National Excellence Program \'UNKP, the NKFIA research grants 123842, 123959, 124845, 124850 and 125105 (Hungary); the Council of Science and Industrial Research, India; the HOMING PLUS program of the Foundation for Polish Science, cofinanced from European Union, Regional Development Fund, the Mobility Plus program of the Ministry of Science and Higher Education, the National Science Center (Poland), contracts Harmonia 2014/14/M/ST2/00428, Opus 2014/13/B/ST2/02543, 2014/15/B/ST2/03998, and 2015/19/B/ST2/02861, Sonata-bis 2012/07/E/ST2/01406; the National Priorities Research Program by Qatar National Research Fund; the Programa Estatal de Fomento de la Investigaci{\'o}n Cient{\'i}fica y T{\'e}cnica de Excelencia Mar\'{\i}a de Maeztu, grant MDM-2015-0509 and the Programa Severo Ochoa del Principado de Asturias; the Thalis and Aristeia programs cofinanced by EU-ESF and the Greek NSRF; the Rachadapisek Sompot Fund for Postdoctoral Fellowship, Chulalongkorn University and the Chulalongkorn Academic into Its 2nd Century Project Advancement Project (Thailand); the Welch Foundation, contract C-1845; and the Weston Havens Foundation (USA).
\end{acknowledgments}

\bibliography{auto_generated}
\cleardoublepage \appendix\section{The CMS Collaboration \label{app:collab}}\begin{sloppypar}\hyphenpenalty=5000\widowpenalty=500\clubpenalty=5000\vskip\cmsinstskip
\textbf{Yerevan Physics Institute, Yerevan, Armenia}\\*[0pt]
A.M.~Sirunyan, A.~Tumasyan
\vskip\cmsinstskip
\textbf{Institut f\"{u}r Hochenergiephysik, Wien, Austria}\\*[0pt]
W.~Adam, F.~Ambrogi, E.~Asilar, T.~Bergauer, J.~Brandstetter, M.~Dragicevic, J.~Er\"{o}, A.~Escalante~Del~Valle, M.~Flechl, R.~Fr\"{u}hwirth\cmsAuthorMark{1}, V.M.~Ghete, J.~Hrubec, M.~Jeitler\cmsAuthorMark{1}, N.~Krammer, I.~Kr\"{a}tschmer, D.~Liko, T.~Madlener, I.~Mikulec, N.~Rad, H.~Rohringer, J.~Schieck\cmsAuthorMark{1}, R.~Sch\"{o}fbeck, M.~Spanring, D.~Spitzbart, A.~Taurok, W.~Waltenberger, J.~Wittmann, C.-E.~Wulz\cmsAuthorMark{1}, M.~Zarucki
% ...
\vskip\cmsinstskip
% ...
1:  Also at Vienna University of Technology, Vienna, Austria\\
% ...
\end{sloppypar}
\end{document}